\documentclass[pre,aps,12pt]{revtex4}
\usepackage{amssymb}
\usepackage{graphicx}
\begin{document}

\title{Magnetic properties of ferro-antiferromagnetic spin triangle chain}
\author{D.~V.~Dmitriev}
\email{dmitriev@deom.chph.ras.ru}
\author{V.~Ya.~Krivnov}
\affiliation{Institute of Biochemical Physics of RAS, Kosygin str.
4, 119334, Moscow, Russia.}
\date{}

\begin{abstract}
We study the frustrated spin-$\frac{1}{2}$ model consisting of a
linear chain of triangles with ferro (F)- and antiferromagnetic
(AF) interactions connected by ferromagnetic interactions
(triangles chain). The ground state phase diagram as a function of
the frustration parameter (a ratio of F and AF interactions) and
the interaction between triangles consists of the ferromagnetic,
two ferrimagnetic and the singlet phases. We study the magnetic
properties in these phases and analyze the magnetization curves.
We show that there are the magnetization plateau in the singlet
phase and the magnetization jumps in some region of this phase. We
study the low-thermodynamics and its relation to the specific
structure of the excitation spectrum of the triangle chain.
\end{abstract}

\maketitle

\section{Introduction}

\begin{figure}[tbp]
\includegraphics[width=4in,angle=0]{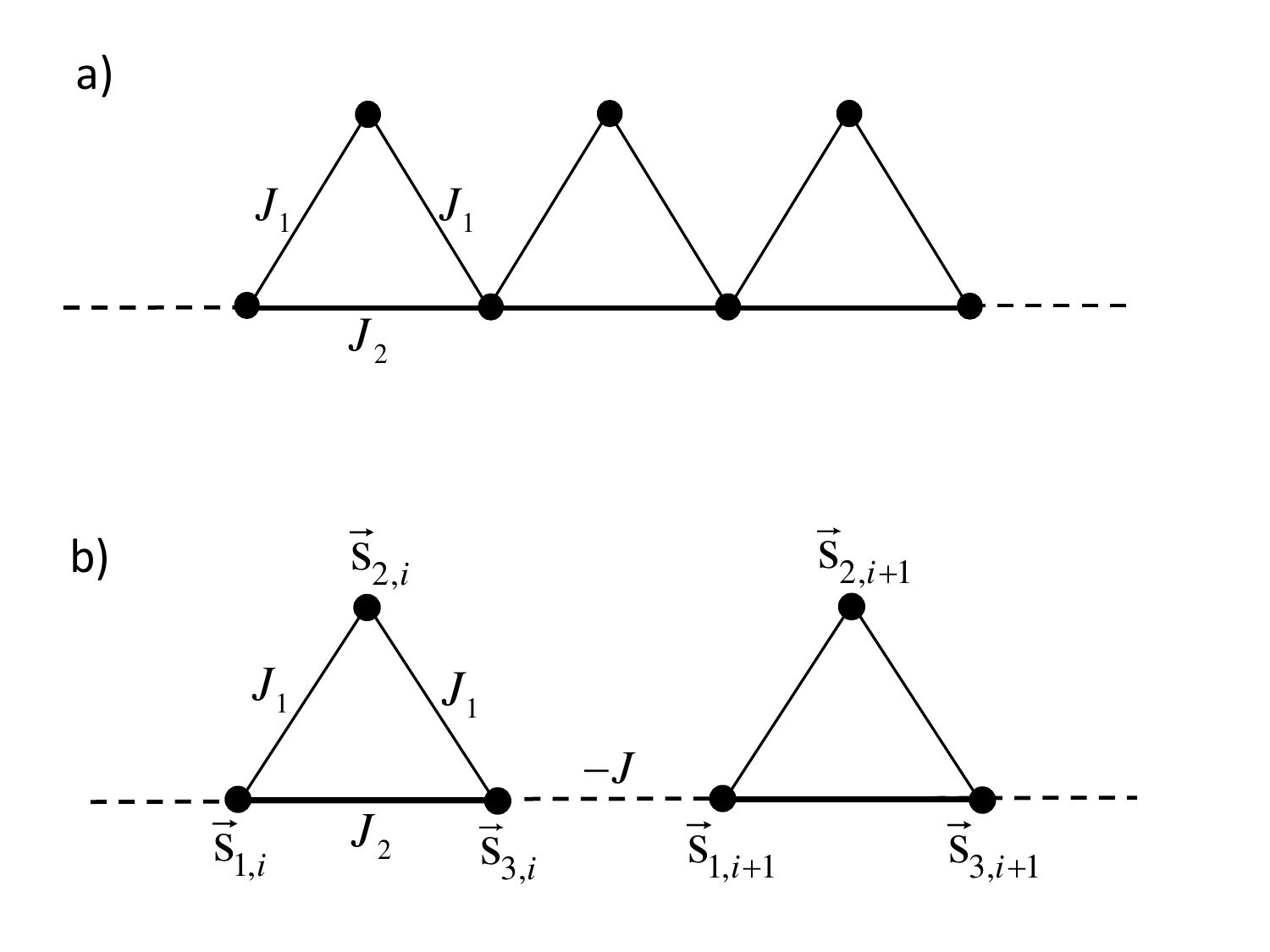}
\caption{(a) $\Delta$-chain model and (b) triangle chain model.}
\label{Fig_triangles}
\end{figure}

Low-dimensional quantum magnetic systems based on geometrically
frustrated lattices are tempting subjects of study both from an
experimental and a theoretical point of view \cite{Diep, Lacrose,
Moesner}. An interesting class of such objects are compounds
consisting of triangular clusters of magnetic ions. These, for
example, include magnets on two-dimensional and
quasi-one-dimensional kagome lattices, and pyrochlore lattices.
The simplest and typical example of such system is
spin-$\frac{1}{2}$ delta-chain (or saw-tooth chain), which is a
linear chain of connected triangles as shown in
Fig.\ref{Fig_triangles}a. The delta-chain with both isotropic
antiferromagnetic interactions $J_{1}>0$ and $J_{2}>0$ is well
studied and has a number of interesting properties \cite{Modern
Physics, Derzhko, Mac, Shulen, Zhit, Zhit2, Capponi, Balika}. At
the same time, the delta-chain with the competing ferro (F) - and
antiferromagnetic (AF) interactions ($J_{1}<0,$ $J_{2}>0$) (the
F-AF delta-chain) is very interesting as well and attract interest
last time \cite{Tonegawa, Kaburagi, DK, KD, DKRS, s12, Brenig,
Rausch,ferri,Schnack,Naturforschung}. The F-AF delta-chain
describes real compounds, in particular, malonate-bridged copper
complexes \cite{malonate, malonate1,InorgChem}, kagome fluoride
$Cs_{2}LiTi_{3}F_{12}$ \cite{Ueda} and cyclic complexes
$Fe_{10}Gd_{10}$ \cite{F10}.

We will study another example of the frustrated spin-$\frac{1}{2}$
model with F and AF interactions, which is an extension of the
$s=\frac{1}{2}$ F-AF delta-chain. This model represents a linear
chain of triangles connected by the ferromagnetic Heisenberg
interactions (triangle chain) as shown in
Fig.\ref{Fig_triangles}b. The interaction $J_{1}$ between the
basal and the apical spins of triangle is ferromagnetic and the
interaction $J_{2}$ is antiferromagnetic. The interaction between
triangles, $-J$, is ferromagnetic. The Hamiltonian of this model
has a form
\begin{equation}
\hat{H}=\sum_{i=1}^{n}\hat{H}_{i}+\sum_{i=1}^{n}\hat{V}_{i,i+1}  \label{H}
\end{equation}%
where%
\begin{eqnarray}
\hat{H}_{i} &=&J_{1}\mathbf{s}_{1,i}\cdot \mathbf{s}_{2,i}+J_{1}\mathbf{s}%
_{2,i}\cdot \mathbf{s}_{3,i}+J_{2}\mathbf{s}_{1,i}\cdot \mathbf{s}_{3,i} \\
\hat{V}_{i,i+1} &=&-J\mathbf{s}_{3,i}\cdot \mathbf{s}_{1,i+1}
\end{eqnarray}%
where $\hat{H}_{i}$ is the Hamiltonian of the $i$-th triangle,
$\hat{V}_{i,i+1}$ is the triangle-triangle interaction and $n$ is
a number of triangles. The periodic boundary conditions are
imposed. Further we put $J_{1}=-1$ and $J_{2}=\alpha $, which
fixes the energy scale by $\left\vert J_{1}\right\vert $.

We focus our study on the influence of the triangle-triangle ferromagnetic
interactions on the ground state properties and the low-temperature
thermodynamics of the triangle chain.

The ground state of model (\ref{H}) is ferromagnetic at $\alpha
<\frac{1}{2}$ and any $J>0$. The value $\alpha _{c}=\frac{1}{2}$
at any $J>0$ corresponds to the quantum critical point separating
the ground state phases. The properties of the triangle chain in
the critical point are highly nontrivial and have been studied in
\cite{a12}. In particular, the ground state is macroscopically
degenerated and the degeneracy of the ground states does not
depend on $J$. However, the spectrum of excitations at $\alpha
=\alpha _{c}$ has a specific structure depending on $J$ and this
structure determines the magnetic properties and the
low-temperature thermodynamics \cite{a12}.

One of the interesting points concerning this model is a nature of
the ground states for $\alpha >\alpha _{c}$ or $\gamma >0$
($\gamma =\alpha -\frac{1}{2}$) and its dependence on $J$. At
$J=0$\ the model consists of independent triangles with a doublet
ground state and the ground state of $n$ triangles is $2^{n}$
-fold degenerated. In the limit $J\to \infty $ the basal spins
$\frac{1}{2}$ connected by the ferromagnetic interaction form
triplet, and the $s=\frac{1}{2}$ triangle chain reduces to the
F-AF delta chain with basal spins $s=1$ and apical spins
$s=\frac{1}{2}$. This F-AF delta-chain has been studied in
\cite{nishimoto, s1}. It was shown that this model has five ground
state phases including the ferromagnetic, two ferrimagnetic and
two singlet phases for definite values of the basal-basal
interaction $\alpha $.

Generally, the ground state of the $s=\frac{1}{2}$ triangle chain
depends on frustration parameter, $\alpha$, and interaction
between triangles, $J$. We will show that this model for $J\ll
\gamma $ is described by the effective Hamiltonian, which governs
the ground state properties and low-energy excitations. In order
to study the ground state and the spectrum of the model in the
region of parameters $J\geq \gamma $ we employ exact
diagonalization (ED) and DMRG calculations.

The paper is organized as follows. In Section II we consider the
magnetic properties of the triangle chain in ferrimagnetic phases.
We study the magnetization curve $m(h)$ and the specific heat
$C(T)$ and show that properties of the temperature dependence of
$C(T)$ are related to the specific structure of the excitation
spectrum. In Section III we consider the model in the singlet
phase. We show that in the parametric regime $J\ll \gamma $ the
triangle chain is described by the effective Heisenberg and XXZ
Hamiltonians. We also study the relation of the excitations
spectrum to the behavior of $C(T)$. We show that properties of the
model at large $\alpha$ in the limits $J\to 0$ and $J\to \infty $
are similar to each other. The magnetization plateau at
$m=\frac{1}{2}$ exists in most of the singlet phase excluding the
region close to the border with Ferri I phase. In this region
jumps of the magnetization appear. In Section IV we give a summary
of our results.

\section{Ferrimagnetic phases}

As was noted above the ground state is ferromagnetic at $\alpha
<\frac{1}{2}$ and the critical point $\alpha _{c}=\frac{1}{2}$
separates the ferromagnetic phase from other ones. In the limit
$J\to \infty $ the triangle model reduces to the delta-chain with
basal spins $s=1$, which has the singlet and two ferrimagnetic
ground state phases at $\alpha
>\frac{1}{2}$ \cite{nishimoto, s1}. In particular, in this limit
the ferrimagnetic phase with $S_{tot}\simeq n$ (Ferri 1) exists in
the region $0.5<\alpha \lesssim 0.66$ \cite{nishimoto}. It can be
expected that this phase exists for finite values $J$ as well, and
there is the phase boundary in ($\alpha ,J$) plane between this
ferrimagnetic and the singlet phases. We determined this phase
boundary on a base of numerical calculations of finite triangle
chains as follows. For the chain with $n$ triangle ($n=4,\ldots
20$) at fixed $J$ we determined the value $\alpha (J,n)$, for
which singlet ground state becomes degenerate with another ground
state with total spin $S^{\ast }$. Then we determine the phase
boundary $\alpha (J)$ using linear fitting function $\alpha
(J,n)=\alpha (J)+c/n$ with fitting parameter $c$. We note that the
ground state total spin $S^{\ast }$ in the ferrimagnetic phase can
be slightly over the value $S^{\ast }=n$. For example, in the
limit $J\to \infty $ it is $S^{\ast }=1.04n$ \cite{s1}.

\begin{figure}[tbp]
\includegraphics[width=4in,angle=0]{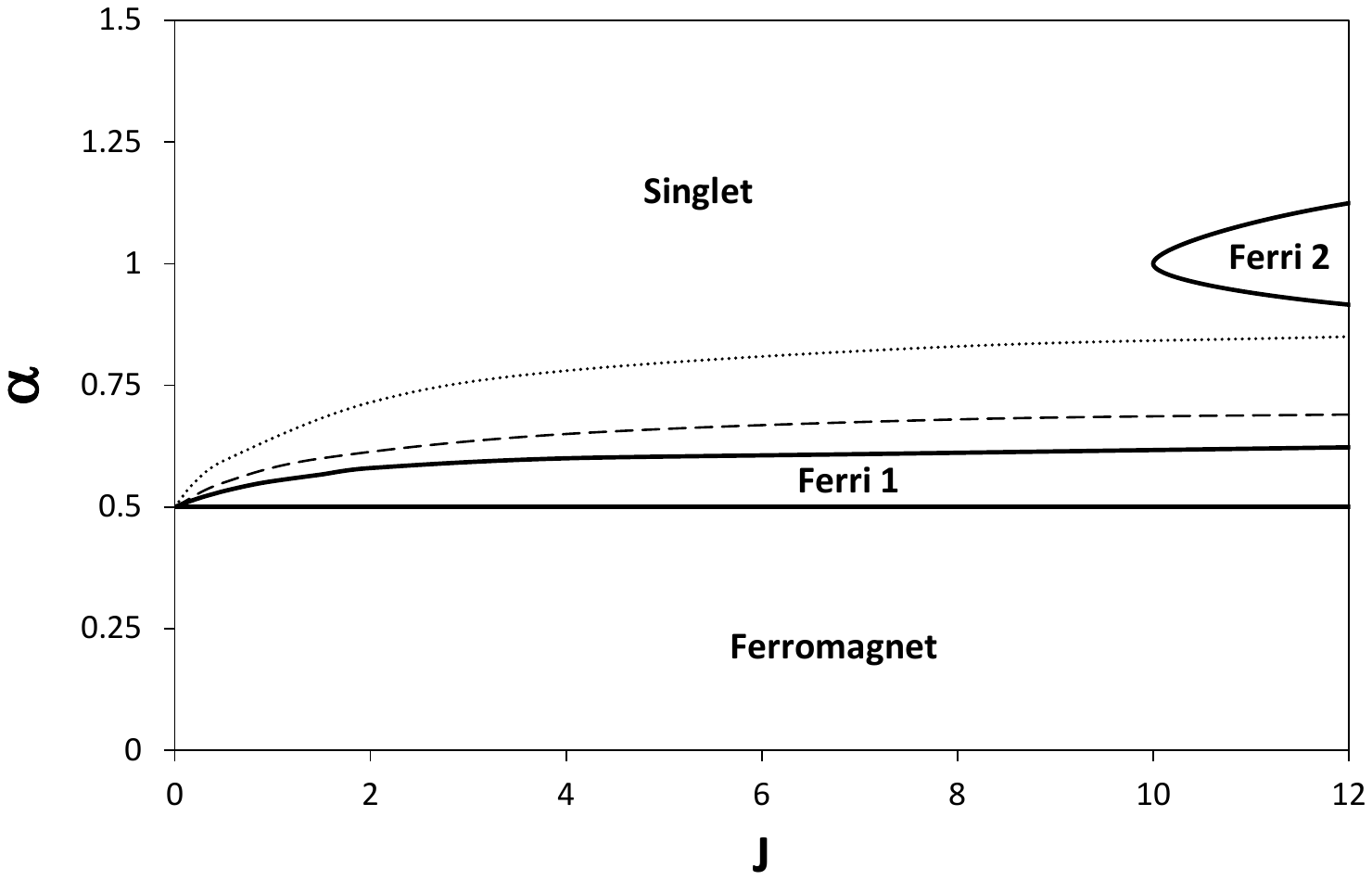}
\caption{Phase diagram of triangle chain model. Dotted line
separates the singlet region without jump on the magnetization
curve (above dotted line) from the singlet region with the
magnetization jump (below dotted line). Dashed line separates the
singlet region with plateau on the magnetization curve (above
dashed line) from the singlet region without the plateau (below
dashed line).} \label{phase}
\end{figure}

Similarly, we define the phase boundary between the singlet and
another ferrimagnetic phase with $S_{tot}=\frac{n}{2}$ (Ferri 2),
which is an extension of the ferrimagnetic phase existing in the
limit $J\to \infty $ in region $0.86\lesssim \alpha \lesssim 1.2$.
The final ground state phase diagram of the triangle chain is
shown in Fig.\ref{phase}. It consists of the ferromagnetic, the
singlet and two ferrimagnetic phases, Ferri 1 and Ferri 2.

We begin the discussion of the properties of the ferrimagnetic
phase Ferri 1 with the ground state magnetization curve. The
magnetization per triangle $m(h)$ at $T=0$ increases monotonically
from $m\simeq 1$ at $h=0$ to $m=\frac{3}{2}$ at $h\to h_{s}$ (the
saturation field is $h_{s}=\gamma $ for any values of $J>0$). (We
use further the dimensionless magnetic field $h=g\mu
_{B}B/\left\vert J_{1}\right\vert $ and the dimensionless
temperature scaled by $\left\vert J_{1}\right\vert $). In the
vicinity of $h_{s}$ the magnetization $m(h)$ is
\begin{equation}
m=\frac{3}{2}-\frac{2\sqrt{2}}{\pi \sqrt{3-\gamma }}\sqrt{1-\frac{h}{\gamma }%
}
\end{equation}

\begin{figure}[tbp]
\includegraphics[width=4in,angle=0]{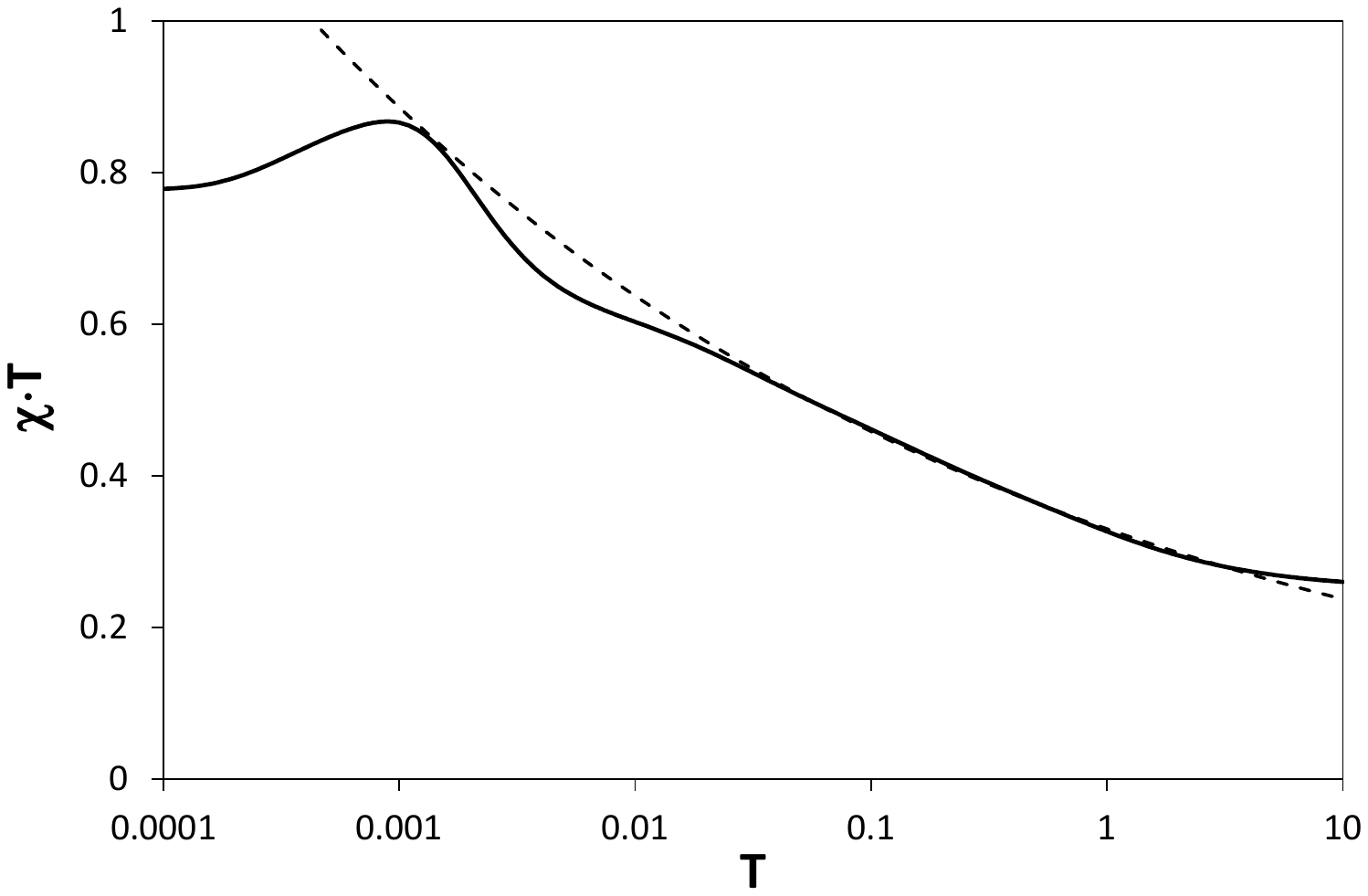}
\caption{Dependence of magnetic susceptibility on temperature for
$\alpha=0.501$ and $J=0.1$, calculated for $n=6$. Dashed line
denotes power-law fit.} \label{chi_T_ferri}
\end{figure}

The zero-field susceptibility $\chi _{0}(T)$ per site diverges at
$T\to 0$, but the quantity $\chi _{0}T$, according to numerical
calculations, is finite at $T=0$ and $\chi
_{0}T=\frac{S_{tot}(S_{tot}+1)}{3n}\simeq \frac{n}{3}$, i.e. the
susceptibility per site is $\chi _{0}\sim \frac{n}{T}$. Such
non-thermodynamic behavior means that $\chi _{0}(T)$ at $T\to 0$
for finite chains is described by the scaling function of type
$\chi _{0}(T)=T^{-\mu }f(nT^{\mu -1})$, which leads to a power-low
dependence $\chi _{0}(T)\sim T^{-\mu }$ for $n\to \infty $. An
estimate of the exponent $\mu $ on a base of the numerical
calculations gives the value which is slightly greater than one.
For example, $\mu \simeq 1.14$ for $J=0.1$ and $\gamma =0.001$
(see Fig.\ref{chi_T_ferri}).

\begin{figure}[tbp]
\includegraphics[width=4in,angle=0]{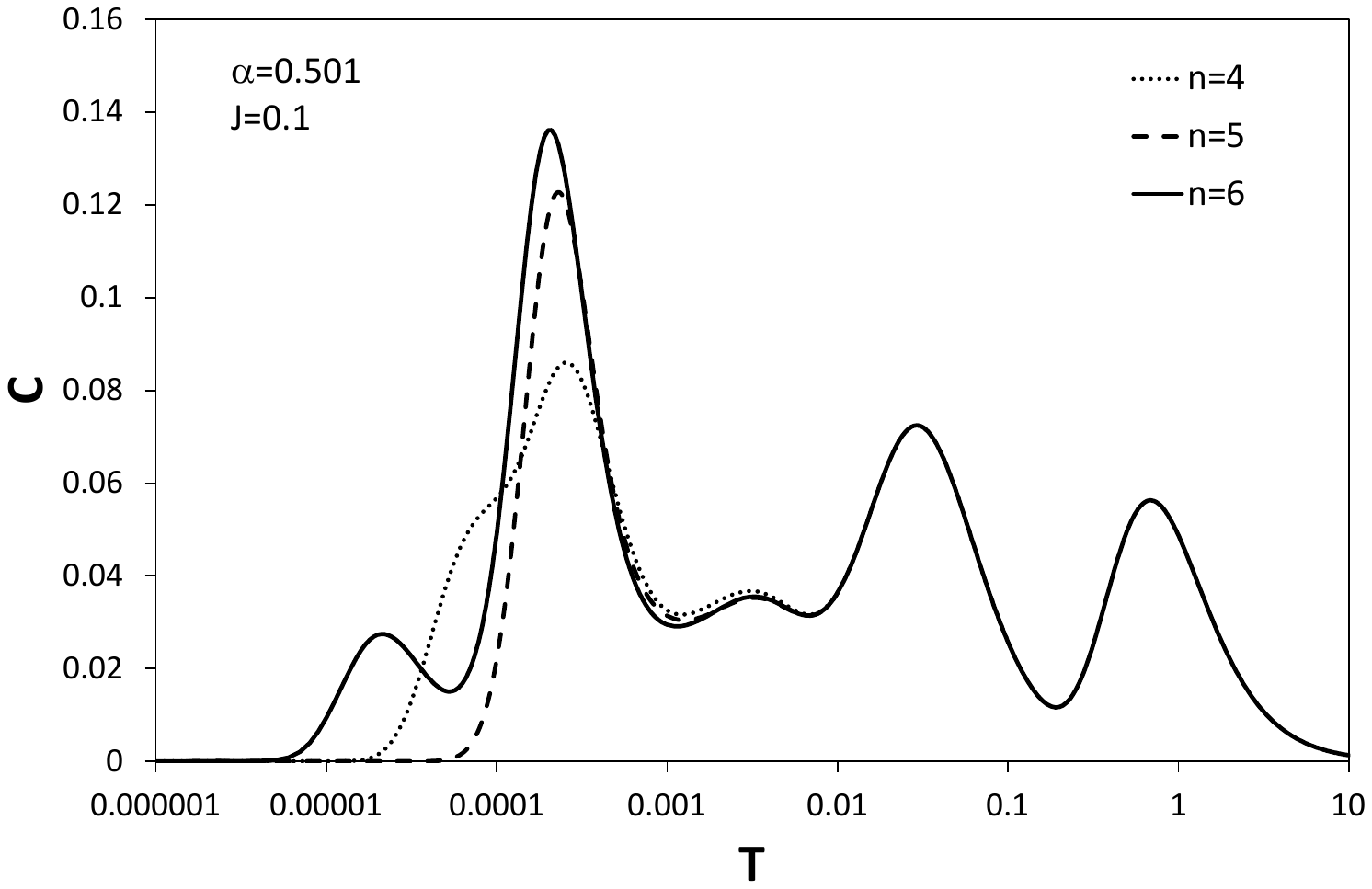}
\caption{Dependence of specific heat on temperature for
$\alpha=0.501 $ and $J=0.1$, calculated for $n=4,5,6$.}
\label{C_T_a0501_J01_L456}
\end{figure}

The behavior of the specific heat in this ferrimagnetic phase is
quite interesting. The temperature dependence $C(T)$ of the
triangle chain with $n=4,5,6$ obtained by the (ED) calculations
for $\gamma =0.001$ and $J=0.1$ is presented in
Fig.\ref{C_T_a0501_J01_L456} (in logarithmic temperature scale).
As it can be seen from Fig.\ref{C_T_a0501_J01_L456} the data for
$n=4,5,6$ practically coincide for $T>10^{-3}$ and the data for
$n=6$ describe the thermodynamic limit at $T>10^{-3}$. As to
positions of the low temperature maximum the data are very close
for different $n$ and the values of the low-temperature maximum
approach finite thermodynamic limit.

\begin{figure}[tbp]
\includegraphics[width=4in,angle=0]{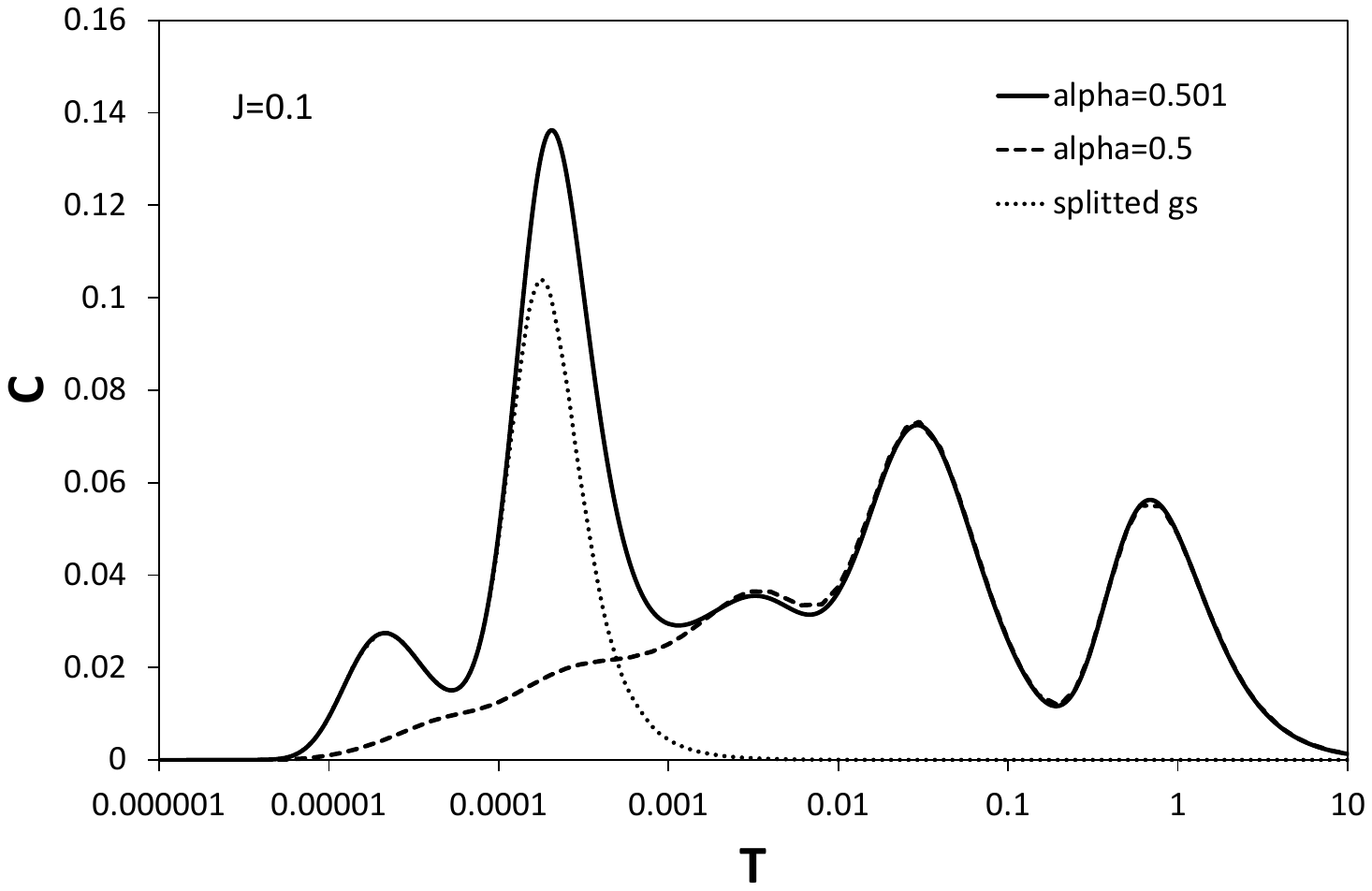}
\caption{Dependence of specific heat on temperature for
$\alpha=0.501 $ and $J=0.1$ compared to that at critical line
$\alpha=0.5$ and $J=0.1$. Dotted line denotes contribution of
split ground states. All curves are calculated for $n=6$.}
\label{C_T_J01_a0501_a05}
\end{figure}

\begin{figure}[tbp]
\includegraphics[width=4in,angle=0]{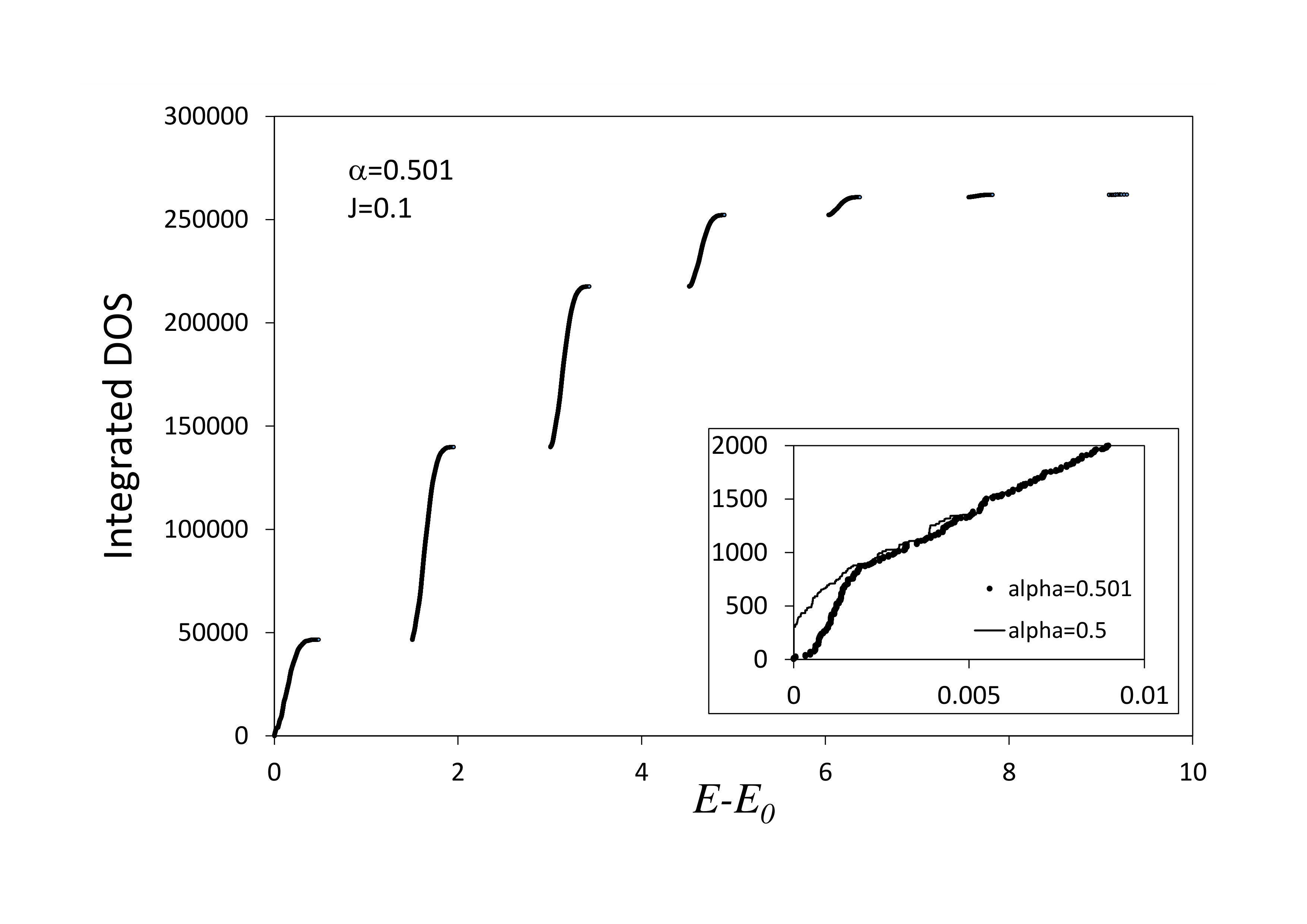}
\caption{Integrated density of states for $\alpha=0.501$ and
$J=0.1$, calculated for $n=6$. The inset shows the lowest energy
part of the plot.} \label{DOS_ferri}
\end{figure}

Together with the low-temperature peak at $T\simeq 10^{-3}$, the specific
heat $C(T)$\ has two maxima at $T\simeq 0.03$ and $T\simeq 1$. Positions and
heights of these two maxima are very close to those for the model in the
critical point, $\gamma _{c}=0$ (see Fig.\ref{C_T_J01_a0501_a05}). As it was
shown in \cite{a12}, their existence for $\gamma =0$ is related to the
step-like structure of the excitation spectrum with gaps between neighboring
bands. The step-like structure of the spectrum has a similar form at $\gamma
>0$, as it is shown in Fig.\ref{DOS_ferri}. The lowest band contains $6^{n}$
states, which determine the broad maximum at $T\simeq 0.03$ and
the maximum at $T\simeq 1$ is related to other high-energy states.
At $\gamma =0$ the lowest band of $6^{n}$ states also contains
$\sim 2^{n}$ degenerated ground states. These states do not give a
contribution to the thermodynamics at $\gamma =0$ because the
partition function of them does not depend on the temperature. But
if $\gamma >0$ these ground states split (as shown in inset of
Fig.\ref{DOS_ferri}) and the low temperature maximum, determined
mainly by them, appears as it is illustrated in
Fig.\ref{C_T_J01_a0501_a05}. This maximum is very narrow but
higher than two others.

There is another ferrimagnetic phase, Ferri 2, with the ground
state spin $S_{tot}=\frac{n}{2}$. This ground state phase is an
extension of the phase existing at $J=\infty $ to a tiny region of
large $J$ as it is shown in Fig.\ref{phase}. The properties of the
triangle chain in this phase are the same as in the delta-chain
with basal spins $s=1$ and apical spins $s=\frac{1}{2}$, studied
in \cite{nishimoto, s1}. In particular, there is the gap for the
state with $S_{tot}=\frac{n}{2}$, which manifests itself as the
magnetization plateau at $m=\frac{1}{2}$. The width of the plateau
equals the gap $\Delta E=E(\frac{n}{2}+1)-E(\frac{n}{2})$, which
is $\Delta E\simeq 0.017$ in this phase. In the vicinity of the
upper field of plateau, at $h\geq h_{c2}$, as well as near the
saturation field $h_{s}=\gamma $ the magnetization has the
square-root singularity.

\section{Singlet phase}

\subsection{Effective Hamiltonians for $J\ll \gamma $}

We begin to study the singlet phase with the case $J\ll \gamma $.
At $J=0$ the system consists of non-interacting triangles. In each
triangle there are eight eigenstates, including four states with
the total spin $S=\frac{3}{2}$ and the energy $\varepsilon
_{0}=-\frac{1}{2}+\frac{\alpha }{4}$ and two doublets. One of them
has the energy $\varepsilon _{0}+\frac{3}{2}$ and the energy of
another doublet is ($\varepsilon _{0}-\gamma $). The latter
doublet with components $\varphi _{1/2}=(\left\vert \downarrow
_{1}\uparrow _{2}\uparrow _{3}\right\rangle -\left\vert \uparrow
_{1}\uparrow _{2}\downarrow _{3}\right\rangle )$
($S_{z}=\frac{1}{2}$) and $\varphi _{-1/2}=(\left\vert \downarrow
_{1}\downarrow _{2}\uparrow _{3}\right\rangle -\left\vert \uparrow
_{1}\downarrow _{2}\downarrow _{3}\right\rangle )$
($S_{z}=-\frac{1}{2}$) is the ground state of the triangle. As a
result, the ground state of total system is $2^{n}$-fold
degenerated with the energy $E_{0}=(\varepsilon _{0}-\gamma )n$.

We note that the triangle chain at $J=0$ can be viewed somewhat
differently. The ground state of this system can be considered as
a product of the basal singlets located on triangles and $n$
decoupled apical spins. The ground state is $2^{n}$ degenerated
and separated by a gap $\Delta E=\gamma $ from the excited states.
This ground state is a direct analogy with the so-called dimer
state of an alternating ferro- antiferromagnetic (F-AF) chain
studied in \cite{Hida, Hung, Sahao}.

The effect of small $J$ can be included within a perturbation
theory. It is convenient to describe two ground states of isolated
triangles as pseudo-spins $\sigma _{i}=\frac{1}{2}$ states, where
$\varphi _{1/2}$ and $\varphi _{-1/2}$ correspond to $\sigma
_{i}^{z}=\frac{1}{2}$ and $\sigma _{i}^{z}=-\frac{1}{2}$. We use
these spin operators for constructing an effective Hamiltonian.
This effective Hamiltonian has a form
\begin{equation}
\hat{H}=\hat{H}_{0}+\hat{V}
\end{equation}%
where $\hat{H}_{0}$ describes the isolated triangles, while
$\hat{V}$ is the interaction between them. The total spin in terms
of the pseudo-spins $\mathbf{\sigma }_{i}$ is $S_{tot}=\sum
\mathbf{\sigma }_{i}$.

It turns out that the first order in $J$ vanishes and the
effective Hamiltonian in the second order in $J$ has a form
\begin{equation}
\hat{H}_{1eff}=E_{0}-nc+J_{eff}\sum_{i}\mathbf{\sigma }_{i}\cdot\mathbf{%
\sigma }_{i+1}  \label{H1}
\end{equation}%
where%
\begin{equation}
c=\frac{3J^{2}(4\gamma ^{2}+7\gamma -1)}{16\gamma (3+2\gamma )(3+4\gamma )}
\end{equation}%
\begin{equation}
J_{eff}=\frac{J^{2}}{4\gamma (3+2\gamma )(3+4\gamma )}
\end{equation}

The Hamiltonian (\ref{H1}) contains a constant and the operator
term. It means that the triangle model in the second order in $J$
consists of the basal ground state singlet and the apical
subsystem. The constant in Eq.(\ref{H1}) is the energy of the
basal singlet and pseudo-spin operators $\mathbf{\sigma }_{i}$ can
be identified with the apical spins $\mathbf{s}_{i}$. The apical
subsystem in the regime $J\ll \gamma $ reduces to the Heisenberg
model with the exchange interaction $J_{eff}$. This model
describes the triangle chain in the spin sectors $0\leq
S_{tot}\leq \frac{n}{2} $, because the maximal value of the total
apical spin is $\frac{n}{2}$.

Let us consider now the triangle chain for the $z$-component of
the total spin $S_{tot}^{z}$ in the range
$\frac{n}{2}<S_{tot}^{z}\leq \frac{3n}{2}$. Again, we begin with
the case $J=0$. In this case the ground state is formed by $(n-k)$
triangles with the state $S^{z}=\frac{3}{2}$ and $k$ triangles
($k=\frac{3n}{2}-S_{tot}^{z}$) with $S^{z}=\frac{1}{2}$ states of
the lowest doublet of the triangle. The ground state energy is
$E(S_{tot}^{z})=n\varepsilon _{0}-k\gamma $ and the number of the
ground states is $C_{n}^{k}=n!/k!/(n-k)! $. For non-zero $J$ (but
$J\ll\gamma$) we can construct another effective Hamiltonian
introducing new pseudo-spin operators $\mathbf{\tau }_{i}$, where
$\tau _{i}^{z}=\frac{1}{2}$ and $\tau _{i}^{z}=-\frac{1}{2}$
correspond to $S^{z}=\frac{3}{2}$ and $\varphi _{1/2}$ states of
the $i$-th triangle, respectively. The total spin is
$S_{tot}^{z}=n+\sum \tau _{i}^{z}$. As a result the effective
Hamiltonian $\hat{H}_{2eff}$ in the first order in $J$ takes a
form
\begin{equation}
\hat{H}_{2eff}=(\varepsilon _{0}-\frac{J}{16}-\frac{\gamma }{2})n-\frac{J}{4}%
\sum \tau _{i}^{z}\tau _{i+1}^{z}+\frac{J}{2}\sum (\tau _{i}^{x}\tau
_{i+1}^{x}+\tau _{i}^{y}\tau _{i+1}^{y})+(\gamma -\frac{J}{4})\sum \tau
_{i}^{z}  \label{H2}
\end{equation}

In the range $\frac{n}{2}<S_{tot}^{z}\leq \frac{3n}{2}$ the
triangle chain reduces to the $XXZ$ model in the magnetic field
$h=(\frac{J}{4}-\gamma )$. Strictly speaking, this $XXZ$ model
describes the ground state of the triangle chain with
$S_{tot}^{z}$, rather than the total spectrum of the system in
this range.

\begin{figure}[tbp]
\includegraphics[width=4in,angle=0]{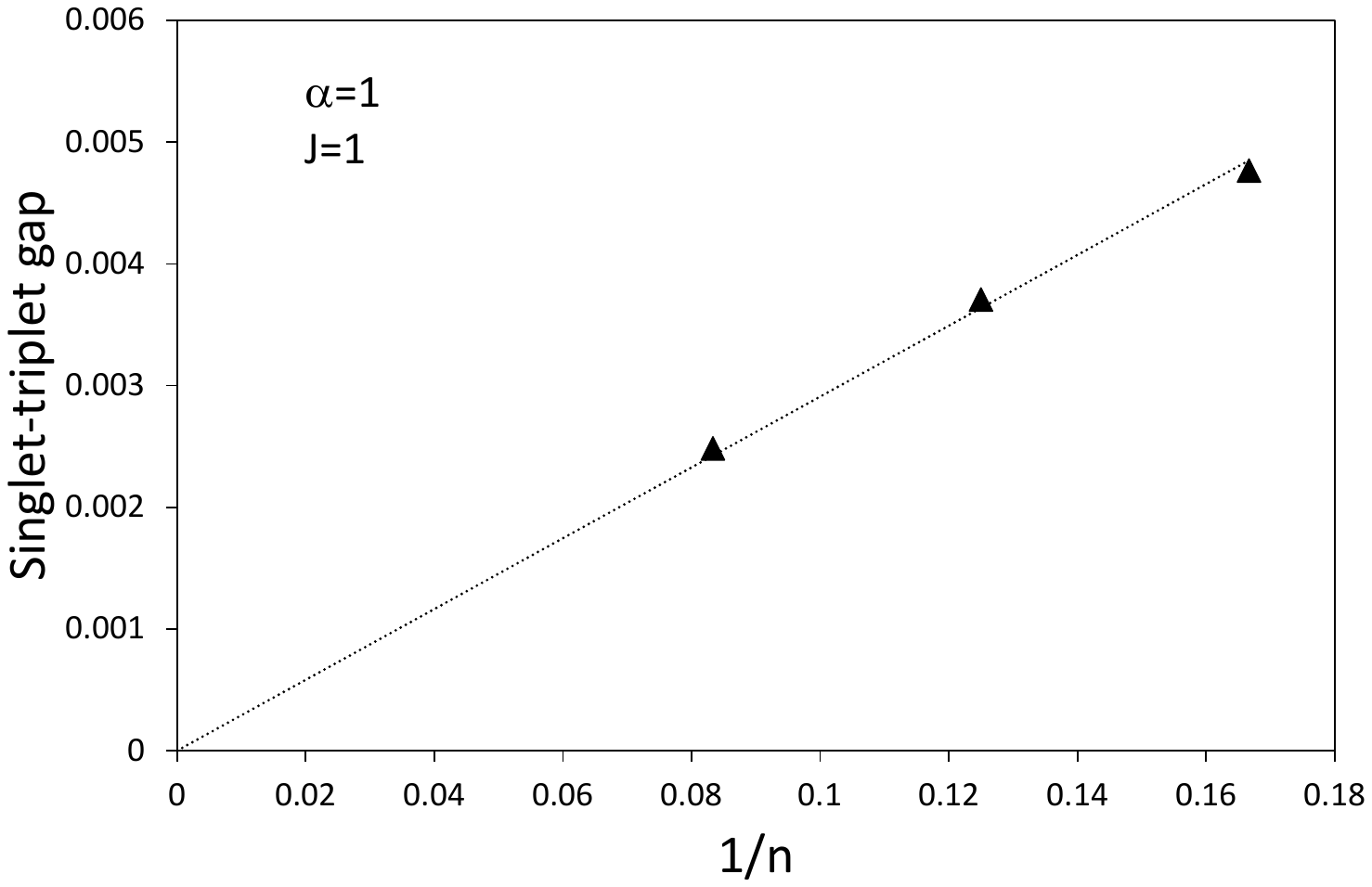}
\caption{Singlet-triplet gap for $\alpha=J=1$ for $n=6,8,12$
plotted vs. $1/n$.} \label{gap_J1_a1}
\end{figure}

In the regime $J\ll\gamma$, the lowest energy in each spin sector
is given by the effective Hamiltonians $\hat{H}_{1eff}$ and
$\hat{H}_{2eff}$. Though the singlet-triplet gap tends to zero at
$n\to \infty $ (see for example Fig.\ref{gap_J1_a1}), there is a
finite gap at $S_{tot}^{z}=\frac{n}{2}$, and this gap is $\Delta
E=E(\frac{n}{2}+1)-E(\frac{n}{2})=\gamma -\frac{J}{2}$ in the
first order in $J$.

The ground state properties and the low-temperature thermodynamics
of the effective Hamiltonians (\ref{H1}) and (\ref{H2}) are well
known \cite{Yang, Griffits} and can be used for the study of those
for the triangle chain. In particular, we can study the dependence
of the ground state energy $E_{0}(S_{tot}^{z})$ and obtain the
magnetization curve $m(h)$ ($m=\frac{S^{z}}{n}$) of the triangle
chain at $J\ll \gamma $. One of the interesting properties of the
ground state magnetization is an existence of the plateau at
$m=\frac{1}{2}$. It is related to the fact that the state with
$S_{tot}^{z}=\frac{n}{2}$ is gapful. The upper and lower magnetic
fields, $h_{c2}$ and $h_{c1}$, determining the width of the
magnetization plateau are $h_{c2}=E(\frac{n}{2}+1)-E(\frac{n}{2})$
and $h_{c1}=E(\frac{n}{2})-E(\frac{n}{2}-1)$. The calculations of
these fields up to the second order in $J$ give
\begin{equation}
h_{c2}=\gamma -\frac{J}{2}-J^{2}[\frac{1}{144\gamma
}+\frac{1}{72(3+2\gamma )}+\frac{5}{18(3+4\gamma )}]  \label{hup}
\end{equation}%
\begin{equation}
h_{c1}=\frac{J^{2}}{2\gamma (3+2\gamma )(3+4\gamma )}  \label{hlow}
\end{equation}

The plateau width is $W=h_{c2}-h_{c1}$.

The magnetization per triangle $m(h)$ behaves as $m\sim h$ at
small $h$ and $m(h)$ has a square-root singularity in the vicinity
of the "saturation field" of $\hat{H}_{1eff}$ which is $h_{c1}$.
In the vicinity of $h_{c1}$
the magnetization is%
\begin{equation}
m=\frac{1}{2}-\frac{2}{\pi \sqrt{J_{eff}}}\sqrt{1-\frac{h}{h_{c1}}}
\end{equation}

Then there is the magnetization plateau between $h_{c1}$ and
$h_{c2}$. The magnetization at $h\gtrsim h_{c2}$ has a square-root
dependence again and $m(h)$ is
\begin{equation}
m=\frac{1}{2}+\frac{2}{\pi }\sqrt{\frac{\gamma -\frac{J}{2}}{J}}\sqrt{1-%
\frac{h}{h_{c2}}}
\end{equation}

At last, in the vicinity of the saturation field of the triangle
chain $h_{s}=\gamma $, the magnetization $m(h)$ is%
\begin{equation}
m=\frac{3}{2}-\frac{2\sqrt{\gamma }}{\pi \sqrt{J}}\sqrt{1-\frac{h}{h_{s}}}
\end{equation}

\begin{figure}[tbp]
\includegraphics[width=4in,angle=0]{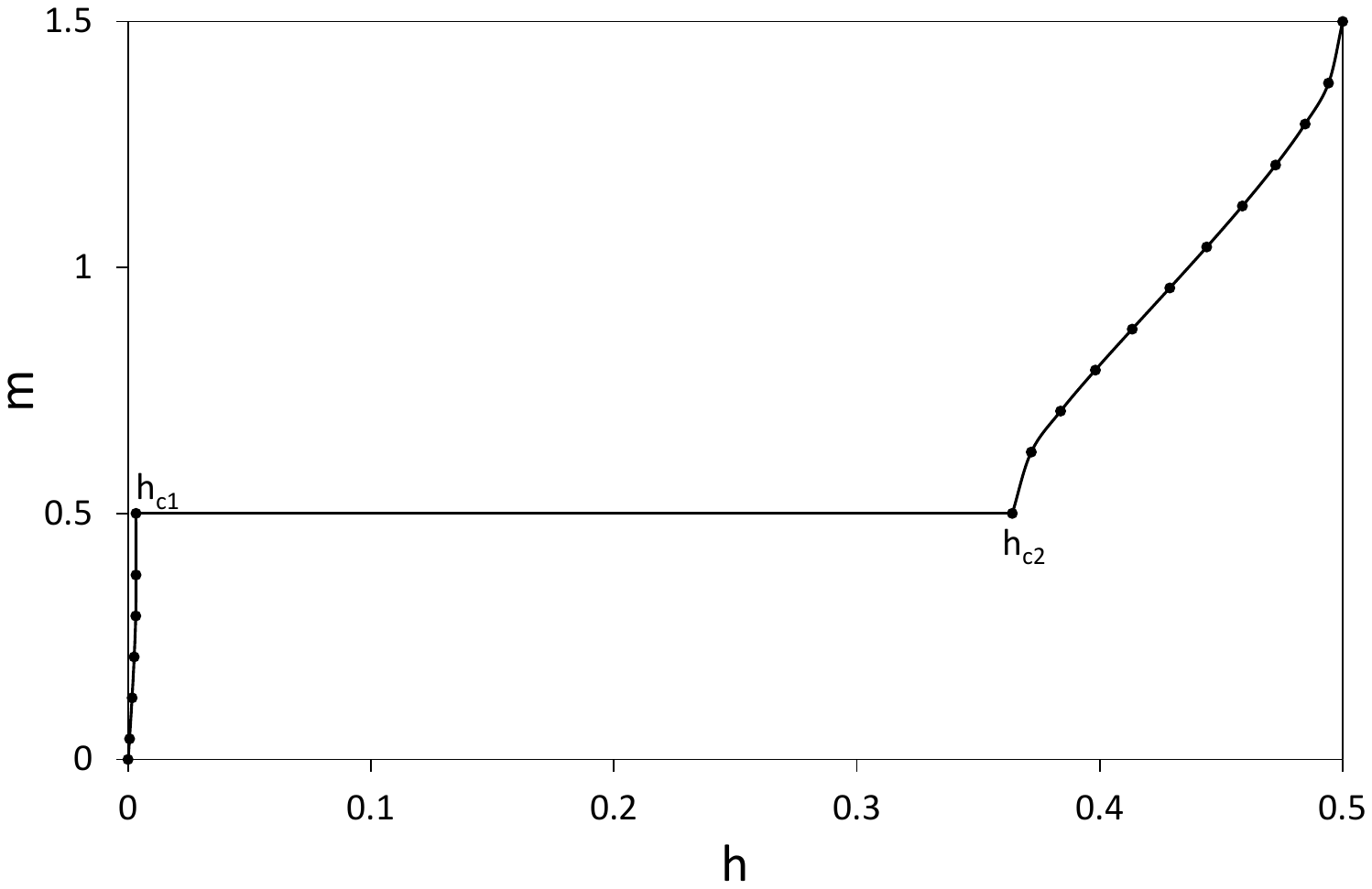}
\caption{Magnetization curve for $\alpha=1$ and $J=0.3$,
calculated for $n=12$.} \label{M_h_a1_J1}
\end{figure}

Numerical calculations of the magnetization of finite triangle
chain agree well with that obtained from the Hamiltonians
(\ref{H1}) and (\ref{H2}) as it is shown in Fig.\ref{M_h_a1_J1}.

\begin{figure}[tbp]
\includegraphics[width=4in,angle=0]{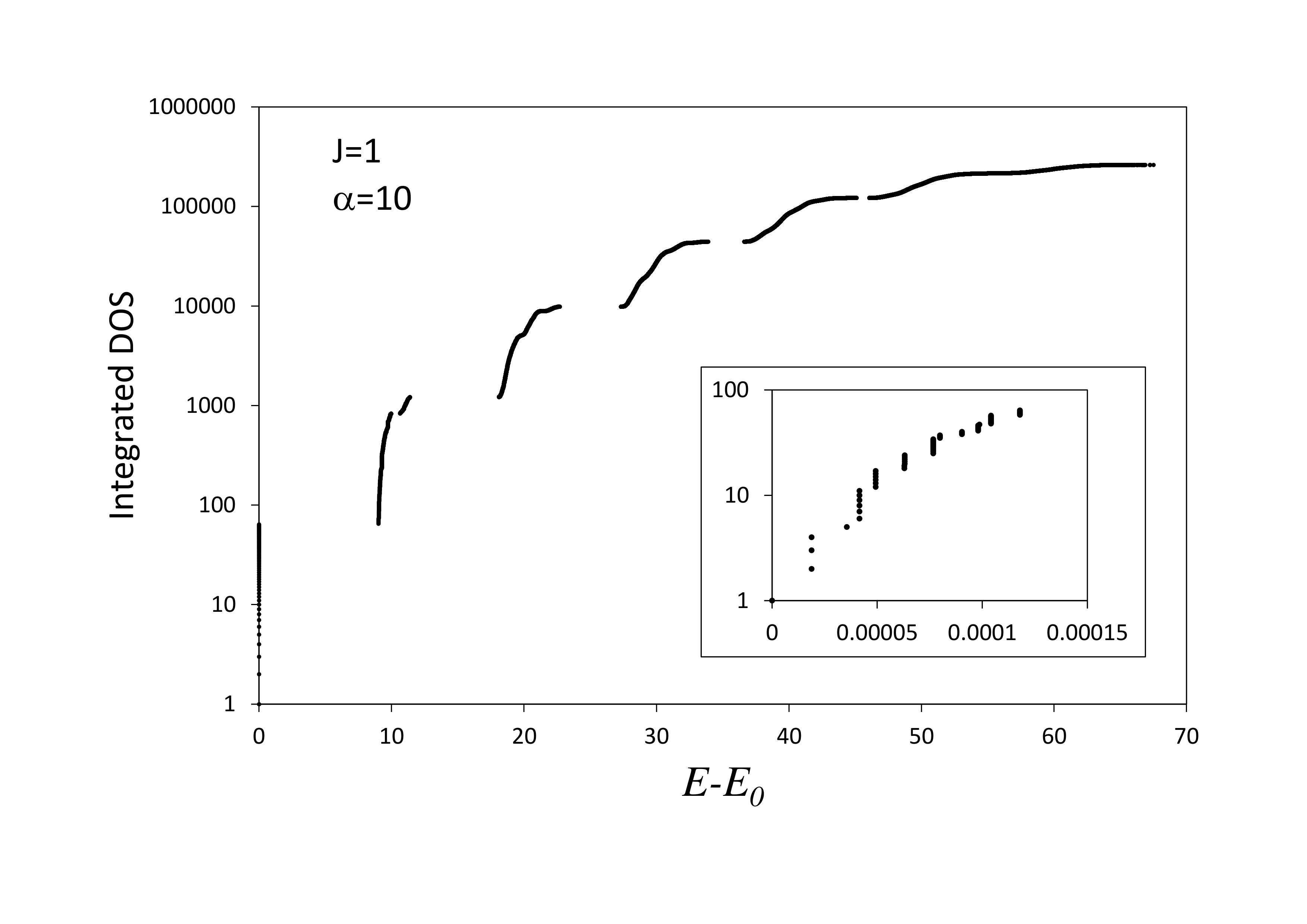}
\caption{Integrated density of states for $\alpha =10$ and $J=1$,
calculated for $n=6$. The inset shows the lowest energy part
containing $2^6$ states.} \label{DOS}
\end{figure}

\begin{figure}[tbp]
\includegraphics[width=4in,angle=0]{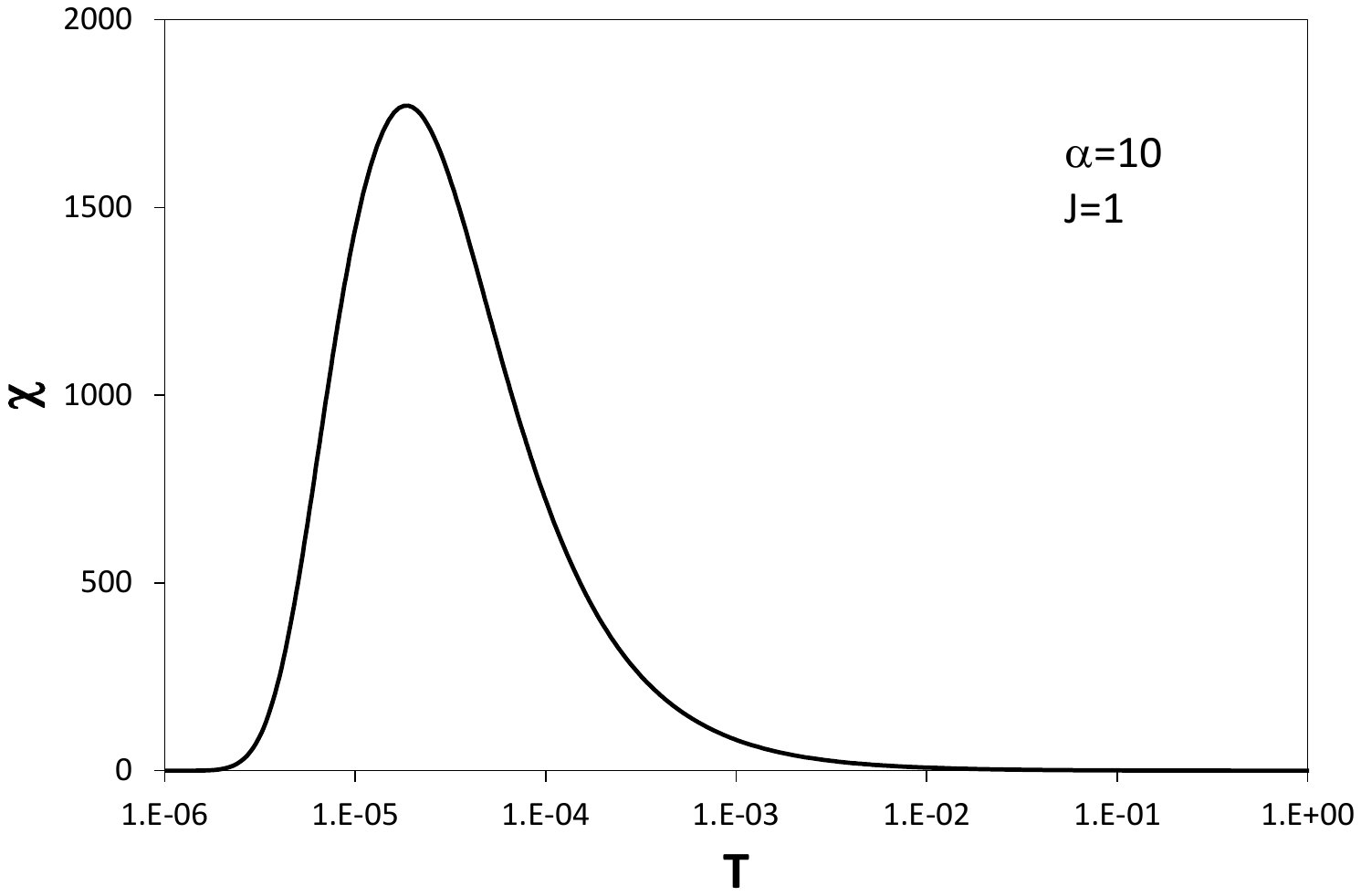}
\caption{Dependence of magnetic susceptibility on temperature for
$\alpha=10$ and $J=1$, calculated for $n=6$.} \label{chi_T_a10_J1}
\end{figure}

As we noted before the lowest energy of the triangle chain for
each $S_{tot}^{z}$ in the limit $J\ll \gamma $ is given by the
effective Hamiltonian. It is interesting to compare the spectrum
of excitations of the effective Hamiltonian with that for the
triangle chain. The integrated density of states of
$\hat{H}_{eff}$ consists of bands separated by gaps $\Delta E\sim
\gamma $. The lowest band contains all $2^{n}$ states of
$H_{1eff}$, and higher bands include the states of
$\hat{H}_{2eff}$. In Fig.\ref{DOS} we present the integrated
density of states of the triangle chain with $n=6$ and $\alpha
=10$ and $J=1$. As it is seen from Fig.\ref{DOS} the spectrum has
a step-like structure as for $\hat{H}_{eff}$ and the lowest band
(shown in inset in Fig.\ref{DOS}) containing $2^{6}$ states,
coincides with that for $H_{1eff}$, but other bands contain more
high-energy states than those for $\hat{H}_{2eff}$, so that the
states of $\hat{H}_{2eff}$ give a very small contributions to
these bands. However, for $T\leq J_{eff}$ the thermodynamics is
determined by the lowest band, and we can use well-known results
obtained for the $s=\frac{1}{2}$ Heisenberg chain for the study of
the low-temperature thermodynamics of the triangle chain. For
example, the zero-field susceptibility per triangle $\chi _{0}(T)$
in the thermodynamic limit $n\to \infty $ has a maximum $\chi
_{\max }=\frac{0.147}{J_{eff}}$ at $T_{\max }=0.64J_{eff}$
\cite{Eggert}. The results of numerical (ED) calculations of $\chi
(T)$ for the triangle chain with $n=6$ and $\alpha =10$ and $J=1$
shown in Fig.\ref{chi_T_a10_J1}. The value $\chi _{\max }$ agrees
within $5\%$ with given above, but $\chi _{0}(T)$ approaches zero
at $T\to 0$ in contrast with the value $\chi _{0}(0)=\frac{1}{\pi
^{2}}$ for $H_{1eff}$ in the thermodynamic limit. This discrepancy
is due to finite-size effects.

\begin{figure}[tbp]
\includegraphics[width=4in,angle=0]{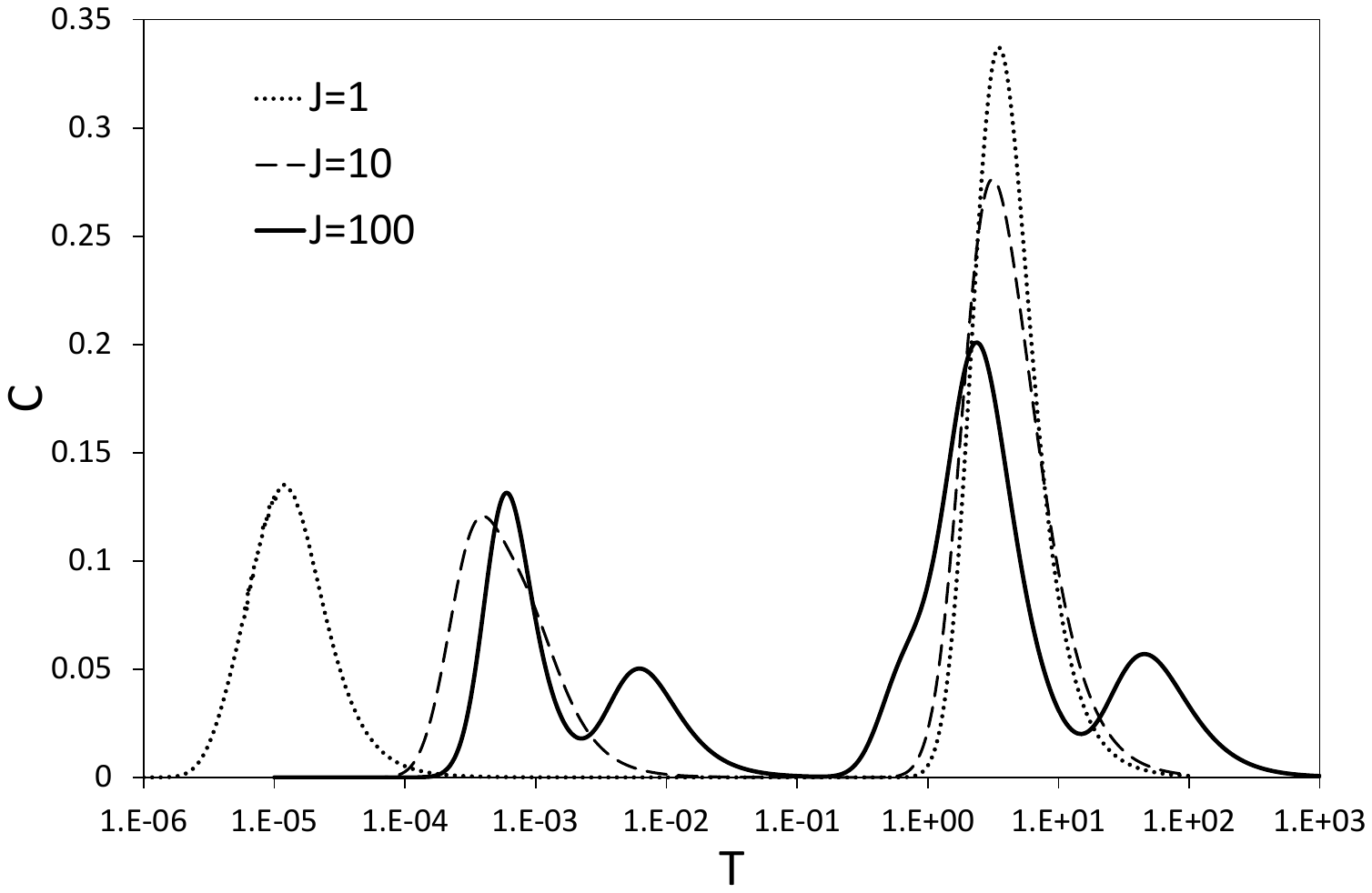}
\caption{Dependence of specific heat on temperature for
$\alpha=10$ and several values of $J=10,30,100$, calculated for
$n=6$.} \label{C_T_a10}
\end{figure}

The specific heat $C(T)$ of the Heisenberg chain has a maximum
$C_{\max }=0.3497$ at $T_{\max }=0.4803J_{eff}$ \cite{Klumper}.
The results of the numerical calculations of $C(T)$ are shown in
Fig.\ref{C_T_a10} and the specific heat has two maxima. The
low-temperature maximum, which is determined by $2^{n}$ states of
the lowest band of the spectrum, agrees within $10\%$ with given
above. The second maximum at $T\simeq 1$ is related to all other
states of the triangle chain. The difference between $T_{1\max } $
and $T_{2\max }$ approximately equals the energy gap in the
spectrum.

\subsection{Large $\alpha $ limit}

It is interesting to consider how the properties of the model in
the singlet phase evolute with the parameter $J$, when $J$ changes
from $J=0$ to $J\to \infty $. At first, we consider the case of
large value of $\alpha $. As we noted before, the triangle chain
at $\alpha \gg J$ is formed (for $S_{tot}\leq \frac{n}{2}$) by the
basal triangle singlet and the apical subsystem which described by
the effective Heisenberg Hamiltonian $H_{1eff}$. This description
is qualitatively valid up to the values $J\simeq \alpha $. With
the further increase of $J$ it becomes incorrect, though the
ground state of the model also consists of both basal and apical
singlets. In the limit $J\to \infty $ two basal spins connected by
the ferromagnetic interaction form triplet and the triangle model
reduces to the delta-chain with basal spin $s=1$. For large
$\alpha $ the basal subsystem of this delta-chain is the $s=1$
Heisenberg chain with the singlet ground state, and the apical
subsystem is described by another effective Hamiltonian, which
represents two weakly interacting Heisenberg spin-chains on odd
and even apical sites as shown in \cite{Chandra}. The ground state
of this effective Hamiltonian is the singlet and the lowest energy
in the sector with the total apical spin $S_{a}$ is a smoothly
increasing function of $S_{a}$ up to $S_{a}=\frac{n}{2}$. The
state with the total spin $S_{tot}=\frac{n}{2}$ is gapped and the
gap is $\Delta E=\frac{\alpha\Delta E_{1}}{2}-\frac{1}{2}$, where
$\Delta E_{1}=0.4105$ is the Haldane singlet-triplet gap
\cite{White}. This gap determines the upper magnetic field
$h_{c2}$ of the magnetization plateau, while the lower magnetic
field is $h_{c1}\simeq 0.26/\alpha$ \cite{s1}. The magnetization
curve in the limit $J=\infty $ has been obtained in \cite{s1}.

As follows from the above, the main properties of the model with
large $\alpha $ in both limits $J\to 0$ and $J\to \infty $ are
rather similar with each other. In both limits the low-energy
states consist of two noninteracting subsystems: the basal chain
in the singlet state and the apical subsystem described by the
effective Hamiltonian, which smoothly transforms from $XXX$ chain
at $J\to 0$ to the two weakly interacting AF chains at $J\to\infty
$. This applies also to the spectrum of lowest excitations and the
behavior of the magnetization.

\begin{figure}[tbp]
\includegraphics[width=4in,angle=0]{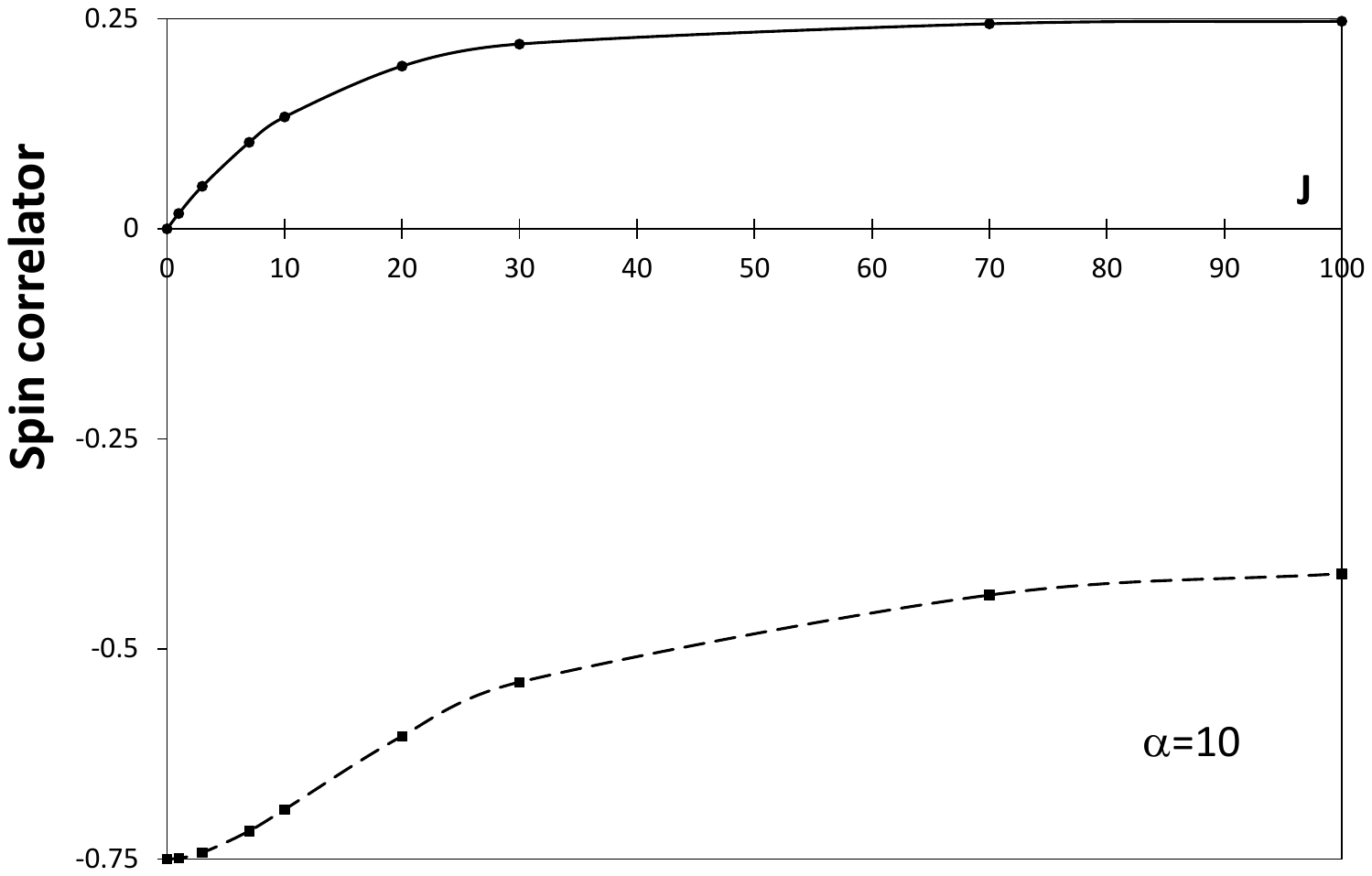}
\caption{Basal local bond strengths on triangle $\left\langle
\mathbf{s}_{1,i}\cdot\mathbf{s}_{3,i}\right\rangle $ (dashed line)
and bond strengths on sites between them $\left\langle
\mathbf{s}_{3,i}\cdot \mathbf{s}_{1,i+1}\right\rangle $ (solid
line) for $\alpha =10$.} \label{S1_S3_a10}
\end{figure}

It is interesting to consider the change of some properties of the
model when the parameter $J$ is changed from $J=0$ to $J\to \infty
$. In Fig.\ref{S1_S3_a10} we show such dependence for basal local
bond strengths on triangle, $\left\langle
\mathbf{s}_{1,i}\cdot\mathbf{s}_{3,i}\right\rangle $, and on sites
between them, $\left\langle \mathbf{s}_{3,i}\cdot
\mathbf{s}_{1,i+1}\right\rangle $, for $\alpha =10$. As it can be
seen from Fig.\ref{S1_S3_a10}, the values of the basal local bond
strengths change continuously from the values at $J=0$ to those at
$J\to \infty $, and the values in these limiting cases agree well
with the corresponding known results. When $J\ll\gamma$ these
correlators are
\begin{eqnarray}
\left\langle \mathbf{s}_{1,i}\cdot\mathbf{s}_{3,i}\right\rangle
&=&-\frac{3}{4}+\frac{J^{2}}{24}[\frac{1}{\gamma
^{2}}+\frac{1}{(3+2\gamma )^{2}}+\frac{16}{(3+4\gamma )^{2}}] \\
\left\langle \mathbf{s}_{3,i}\cdot \mathbf{s}_{1,i+1}\right\rangle
&=&\frac{3J(4\gamma ^{2}+7\gamma +2)}{8\gamma (3+2\gamma
)(3+4\gamma )}
\end{eqnarray}

\begin{figure}[tbp]
\includegraphics[width=4in,angle=0]{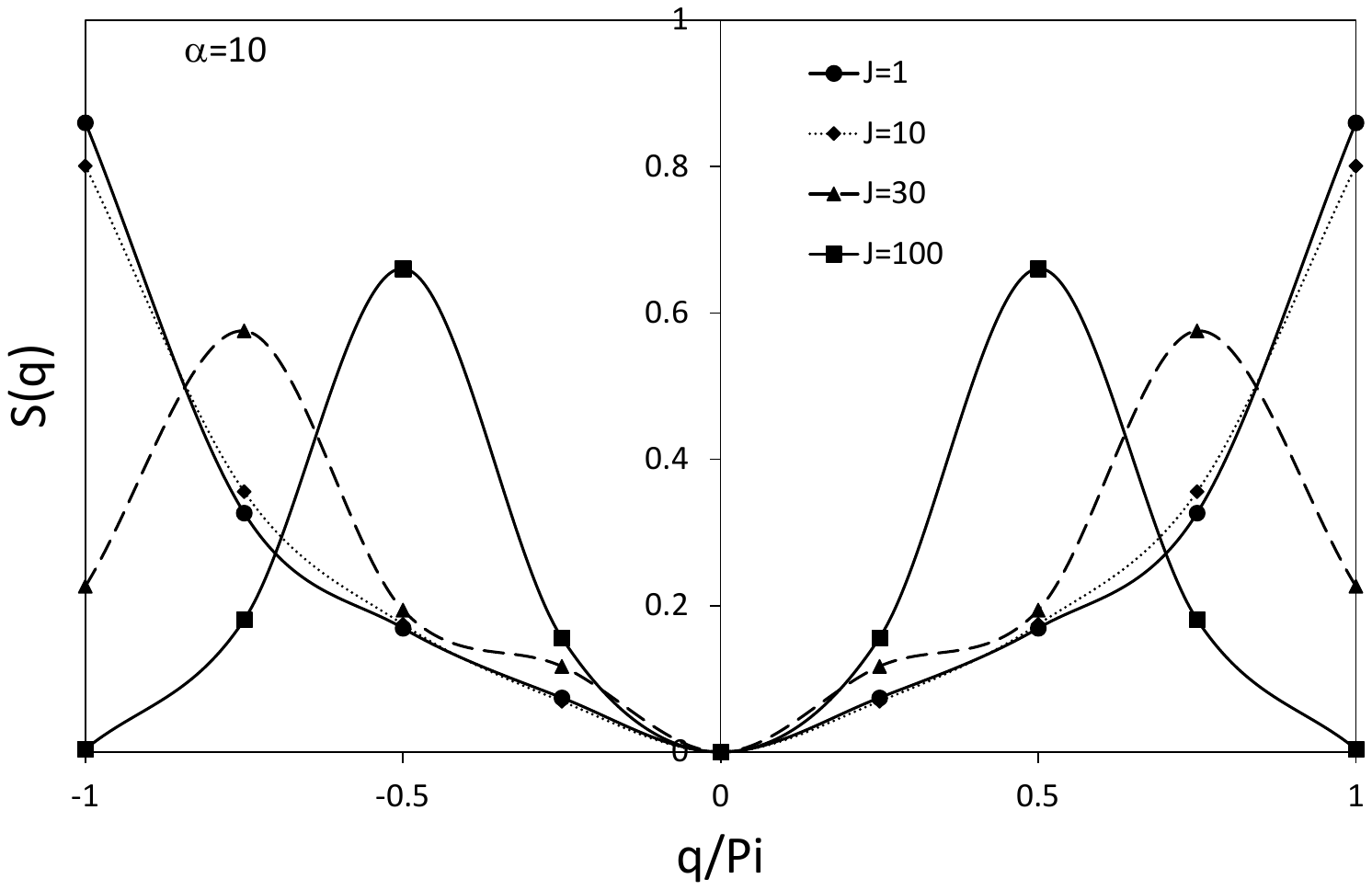}
\caption{Apical spin structure factor for $\alpha =10$ and several
values of $J=1,10,30,100$, calculated for $n=8$.} \label{S_q_a10}
\end{figure}

One of the important quantities of the spin system is the static
structure factor $S(q)$, which characterizes the periodicity of
the system structure. The evolution of the apical spin structure
factor $S(q)$ with changing of the parameter $J$ for $\alpha =10$
and $n=6$ is shown in Fig.\ref{S_q_a10}. As it can be seen from
Fig.\ref{S_q_a10}, the position of the maximum $q_{\max }$ changes
from $q_{\max }=\pi $ in the region $J<\gamma$ to $q_{\max
}=\frac{\pi }{2}$ at $J\gg 1$, taking intermediate values
$\frac{\pi }{2}<q_{\max }<\pi $ in between. This testifies the
incommensurate structure of the triangle chain for intermediate
values of $J$.

The temperature dependence of the specific heat $C(T)$ has two
maxima for different values of $J$ including the cases $J\ll
\alpha $ and $J\gg \alpha $, as shown in Fig.\ref{C_T_a10}. The
difference between the corresponding values of $T_{\max }$ is
approximately equals the gap $\Delta E$ in $S_{tot}=\frac{n}{2}$.
The susceptibility $\chi _{0}(T)$ has a maximum, the temperature
and the height of which continuously change with $J$. The main
conclusion followed from the above consideration is a topological
equivalence of the phase states in two opposite limits $J\to 0$
and $J\to \infty $ and a smooth crossover between these limits of
the model. We note that similar topological equivalence is
observed in the alternating F-AF chain, in which the Haldane phase
adiabatically connected to the dimer phase \cite{Hida}.

\subsection{All interactions of the same order}

\begin{figure}[tbp]
\includegraphics[width=4in,angle=0]{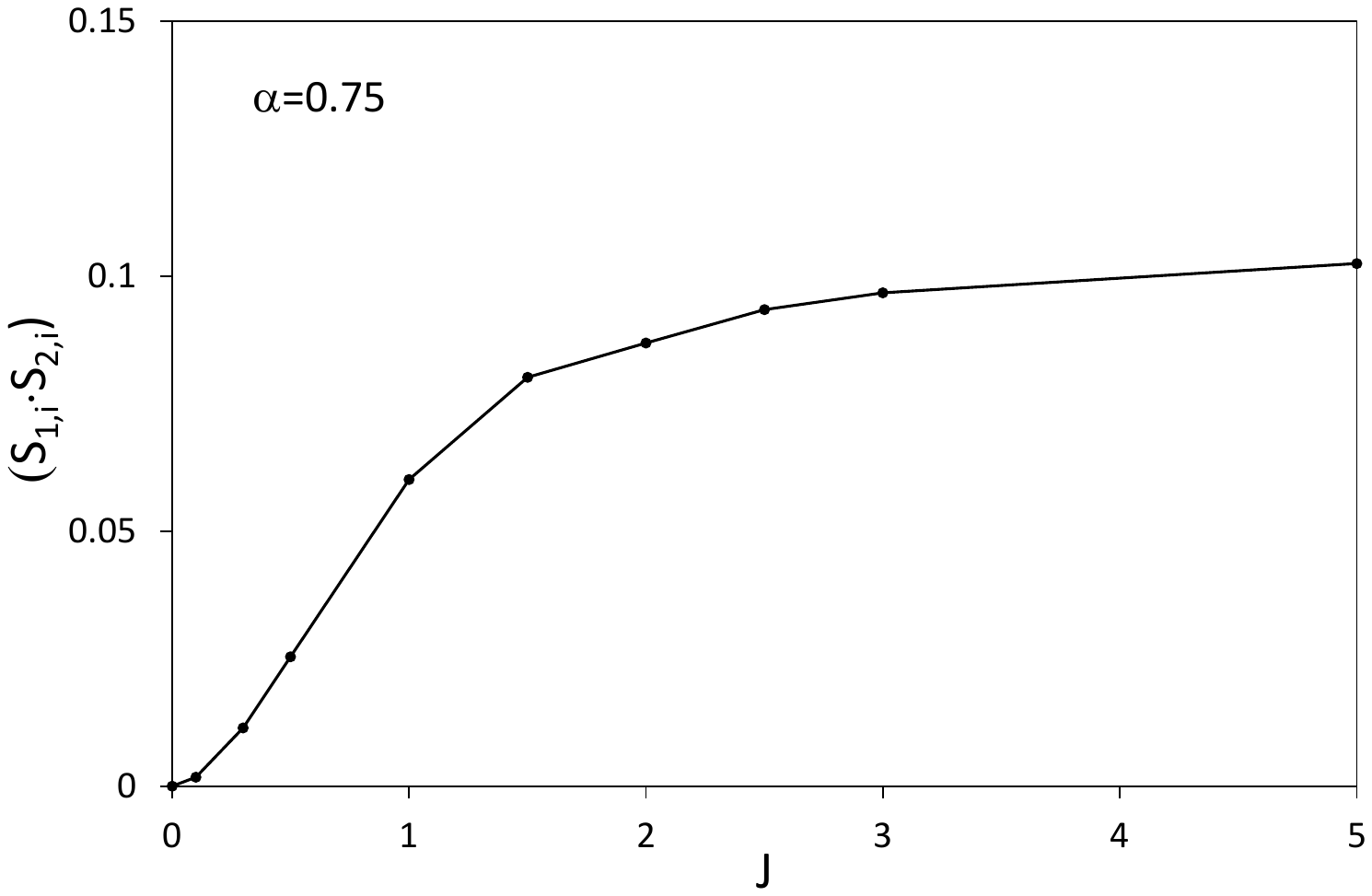}
\caption{Spin correlator between neighboring apical and basal
spins $\left\langle
\mathbf{s}_{1,i}\cdot\mathbf{s}_{2,i}\right\rangle$ as a function
of $J$ for $\alpha=0.75$, calculated for $n=8$.}
\label{S1_S2_a075_L8}
\end{figure}

\begin{figure}[tbp]
\includegraphics[width=4in,angle=0]{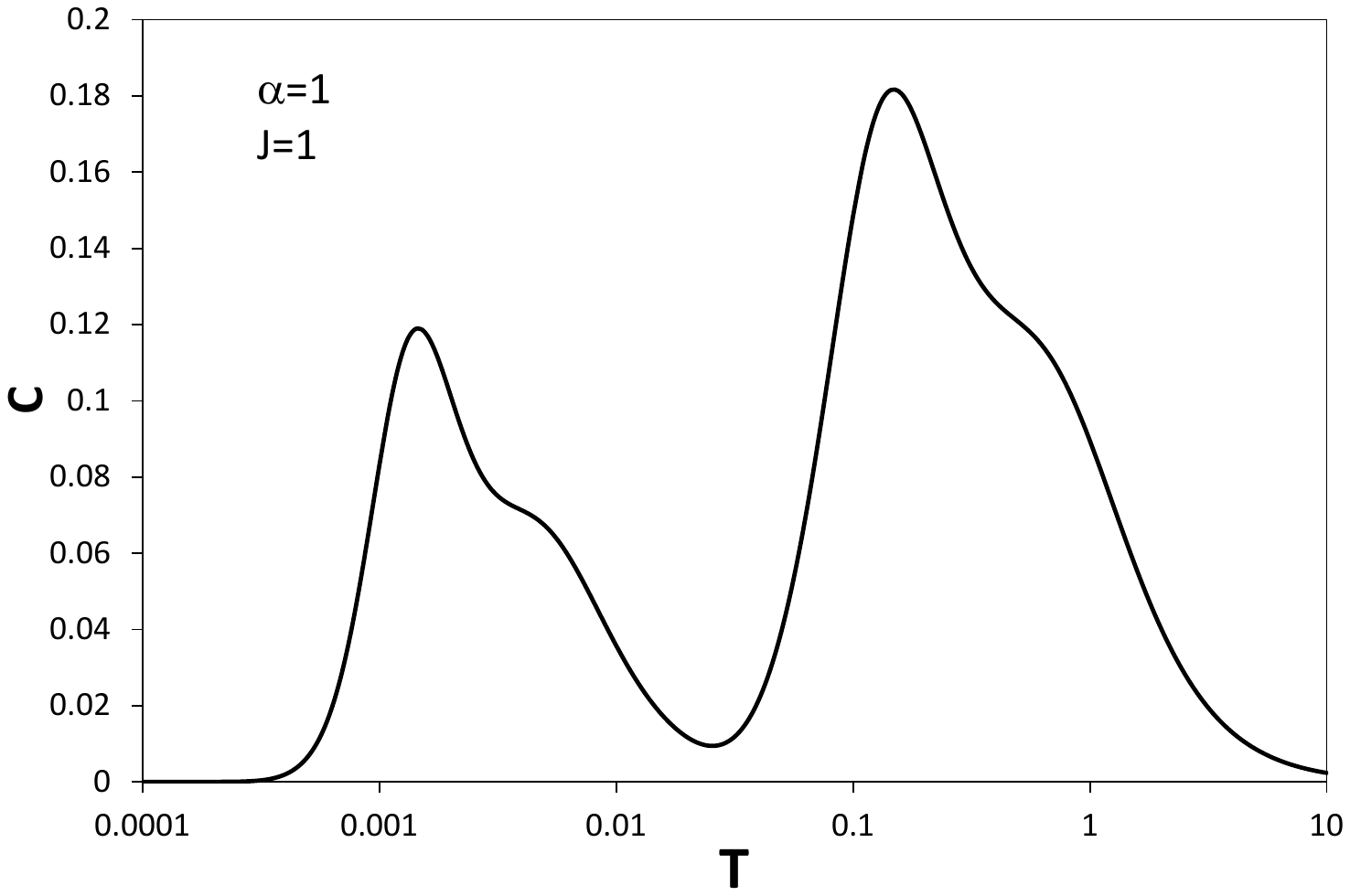}
\caption{Dependence of specific heat on temperature for
$\alpha=J=1$, calculated for $n=6$.} \label{C_T_a1_J1}
\end{figure}

In the part of the phase diagram, where all interactions are of
the same order, the effective separation of the system into almost
independent basal and apical subsystems no longer works. It is
seen in Fig.\ref{S1_S2_a075_L8}, where the spin correlator between
neighboring apical and basal spins is shown as a function of $J$
for $\alpha=0.75$. The correlator is small at small $J$, where the
effective Hamiltonian approach works, but rapidly increases and
takes the finite value when $J\to\infty$. However, it is
surprising that the two-scale energy separation persists even in
this part of the phase diagram. This fact is demonstrated in
Fig.\ref{C_T_a1_J1}, where the dependence of the specific heat on
temperature is shown for $\alpha=J=1$. The dependence $C(T)$
clearly shows two peaks separated by two orders of magnitude in
energy.

\begin{figure}[tbp]
\includegraphics[width=4in,angle=0]{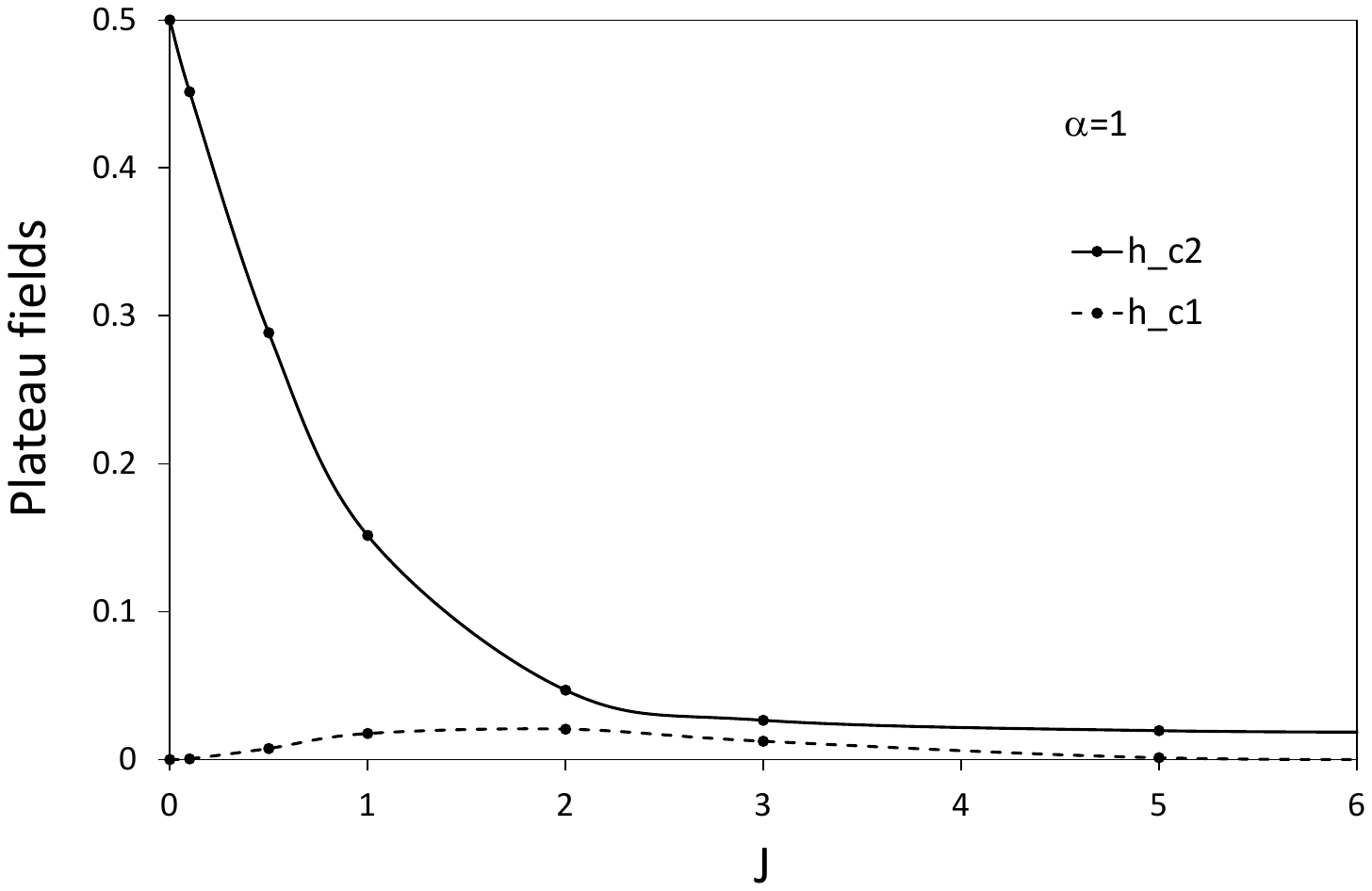}
\caption{Upper ($h_{c2}$) and lower ($h_{c1}$) plateau magnetic
fields as functions of $J$ for $\alpha=1$.} \label{hup_hlow_a1}
\end{figure}

\begin{figure}[tbp]
\includegraphics[width=4in,angle=0]{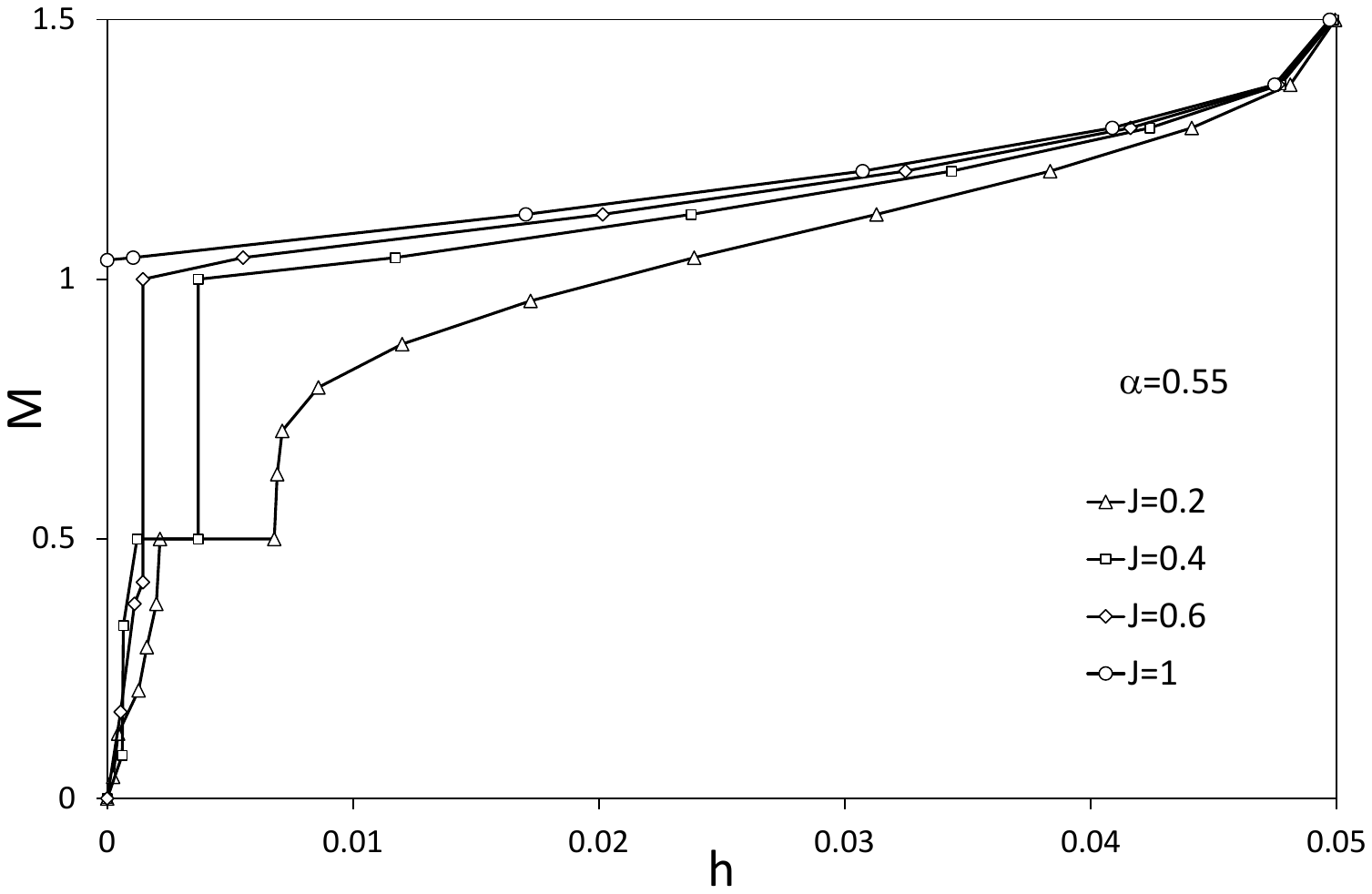}
\caption{Magnetization curves for $\alpha=0.55$ and
$J=0.2;0.4;0.6;1$, calculated for $n=12$. These curves correspond
to different regions on phase diagram Fig.\ref{phase}. The curve
with $J=0.2$ corresponds to singlet phase with magnetization
plateau but without magnetization jump. On the curve with $J=0.4$
the magnetization jump appears, and the magnetization plateau
still exists. The plateau disappears on the curve with $J=0.6$.
The curve with $J=1$ corresponds to Ferri 1 phase.}
\label{M_h_a055}
\end{figure}

The singlet phase in Fig.\ref{phase} is characterized by the
following features. The spectrum of the lowest excitations is
gapless. One of the interesting points related to the model in the
singlet phase is the existence of the gap in the state with
$S_{tot}=\frac{n}{2}$, which leads to the plateau on the
magnetization curve at $m=\frac{1}{2}$. The existence of the
plateau at $m=\frac{1}{2}$ is in accord with the necessary
condition on the magnetization for the appearance of the plateau
given in \cite{plateau}. The lower and upper fields of the plateau
are $h_{c2}=E(\frac{n}{2}+1)-E(\frac{n}{2})$ and
$h_{c1}=E(\frac{n}{2})-E(\frac{n}{2}-1)$. In regime $J\ll \gamma $
the corresponding upper and lower fields of the plateau are given
by Eqs.(\ref{hup}) and (\ref{hlow}). The results for a dependence
of the plateau width (or $h_{c2}$ and $h_{c1}$) on $J$ are
presented in Fig.\ref{hup_hlow_a1} for some representative value
of parameter $\alpha $. The magnetization $m(h)$ increases
continuously from $m=0$ to $m=\frac{1}{2}$ and from
$m=\frac{1}{2}$ to $m=\frac{3}{2}$, when the magnetic field
increases from $h=0$ to $h=h_{c1}$ and from $h=h_{c2}$ to
$h=h_{sat}$, respectively. However, at some value $\gamma _{1}(J)$
(dotted line in Fig.\ref{phase}) close to the border line between
the singlet and Ferri 1 phases the jump of the magnetization at
$m=\frac{1}{2}$ instead of square-root behavior occurs at the
upper field $h_{c2}$. This is a consequence of the dependence of
the energy on $m$, which is downward convex in definite interval
of $m$. With further decrease of $\gamma $ below $\gamma _{1}(J)$
the jump increases, while the width of the plateau decreases and
vanishes on another line $\gamma _{2}(J)$ (dashed line in
Fig.\ref{phase}). At $\gamma <\gamma _{2}(J)$ the magnetization
jump starts at $m_{0}$ and the value $m_{0}$ decreases from
$m_{0}=\frac{1}{2}$ for $\gamma _{2}(J)$ to $m_{0}=0$ on the
border of the Ferri 1 phase, where the jump from $m=0$ to
$m\simeq1$ takes place. In the Ferri 1 phase the jump turns into
the continuous magnetization curve starting from $m\simeq 1$. This
process is illustrated in Fig.\ref{M_h_a055}, where the
magnetization curves presented for $\gamma =0.05$ and different
$J$ for $n=12$.

\section{Summary}

We have studied the ground state and the low-temperature
thermodynamics of the frustrated model, consisting of the
triangles with competing ferro- and antiferromagnetic
interactions. The triangles are connected by the ferromagnetic
Heisenberg interaction. The properties of the model depend on the
frustration parameter $\gamma $ and the triangle-triangle
interaction $J$. The ground state of this triangle chain is
ferromagnetic for $\gamma<0$ at any $J>0$. At the critical point
$\gamma_{c}=0$ the ground state is macroscopically degenerated.
For $\gamma>0$ the ground state phase diagram consists of the
singlet and two ferrimagnetic phases, Ferri 1 and Ferri 2, with
the total spins $S_{tot}\simeq n$ and $S_{tot}=\frac{n}{2}$,
respectively.

The magnetization $m(h)$ in the Ferri 1 phase increases
monotonically from $m\simeq 1$ at $h=0$ to $m=\frac{3}{2}$ at the
saturation field. The zero field susceptibility $\chi _{0}$
diverges at $T\to 0$ as $\chi _{0}\sim T^{-\mu }$ and the exponent
$\mu $ is slightly greater than $1$. The temperature dependence of
the specific heat $C(T)$ has three maxima, which are determined by
the step-like structure of the excitation spectrum. The lowest
energy states split from the macroscopic ground state at $\gamma
=0$ are responsible for the low-temperature maximum arising at
$\gamma >0$, while other states of the lowest band give
contribution to the maximum $C(T)$ at $T\sim 0.03$ and those of
higher band for the maximum at $T\sim 1$.

The ground state spin in the Ferri 2 phase is
$S_{tot}=\frac{n}{2}$ and there is the gap to the state with
$S_{tot}=\frac{n}{2}+1$, which leads to an existence of the
magnetization plateau at $m=\frac{1}{2}$.

The singlet phase is gapless and the state with
$S_{tot}=\frac{n}{2}$ is gapped as for the Ferri 2 phase. The
magnetization plateau exists in the singlet phase, excluding the
narrow region in the vicinity of the Ferri 1 phase. In this
region, as well as in the neighbor small region, the magnetization
jump occurs. This jump turns into the continuous magnetization
curve in the Ferri 1 phase.

The triangle model at large values of $\gamma $ demonstrates a
topological equivalence of the phase states in two opposite limits
$J\to 0$ and $J\to \infty $, and there is a smooth crossover
between these limits of the model.

\begin{acknowledgments}
The numerical calculations were carried out with use of the ALPS
libraries \cite{alps}.
\end{acknowledgments}


\begin{thebibliography}{99}
\bibitem{Diep} H.~T.~Diep (ed) 2013 Frustrated Spin Systems (Singapore;
World Scientific).

\bibitem{Lacrose} C.~Lacroix, P.~Mendels and F.~Mila, eds., Introduction to
frustrated magnetism. Materials, Experiments, Theory(Springer-Verlag,
Berlin, 2011).

\bibitem{Moesner} R.~Moesner, A.~P.~Ramirez, Geometrical frustration,
Physics Today, \textbf{59}, 24 (2006).

\bibitem{Derzhko} O.~Derzhko, J.~Richter, M.~Maksymenko, Int.~J.~Mod.~Phys B
\textbf{29}, 153007 (2015).

\bibitem{Mac} M.\ Maksymenko, A.\ Honecker, R.\ Moessner, J.\ Richter, and
O.\ Derzhko, Phys.\ Rev.\ Lett.\ \textbf{109}, 096404 (2012).

\bibitem{Shulen} J.\ Richter, O.\ Derzhko and J.\ Schulenburg,
Phys.~Rev.~Lett. \textbf{93}, 107206 (2004).

\bibitem{Zhit} M.~E.~Zhitomirsky, H.~Tsunetsugu, Phys.~Rev. B \textbf{70},
100403(R) (2004).

\bibitem{Zhit2} M.\ E.\ Zhitomirsky and H.\ Tsunetsugu, Phys.~Rev. B \textbf{%
75}, 224416 (2007).

\bibitem{Capponi} S.\ Capponi, O.\ Derzhko, A.\ Honecker, A.M.\ Lauchli, J.\
Richter, Phys.\ Rev.\ B \textbf{88}, 144416 (2013).

\bibitem{Balika} J.\ Richter, O.\ Krupnitska, V.\ Balika, T.\ Krokhmalski,
O.\ Derzhko, Phys.\ Rev.\ B \textbf{97}, 024405 (2018)

\bibitem{Modern Physics} O.\ Derzhko, J.\ Richter, M.\ Maksymenko, Int.\ J.\
Mod.\ Phys B \textbf{29}, 153007 (2015).

\bibitem{Rausch} R.Rausch, M.Peschke, C.Plorin, J.Schnack, C.Karrasch,
SciPost.Phys. \textbf{14}, 052 (2023).

\bibitem{Brenig} A.\ Metavitsiadis, C.\ Psaroudaki, W.\ Brenig, Phys.\ Rev.\
B \textbf{101}, 235143 (2020).

\bibitem{DK} V.\ Ya.\ Krivnov, D.\ V.\ Dmitriev, S.\ Nishimoto, S.-L.\
Drechsler, and J.\ Richter, Phys. Rev. B \textbf{90}, 014441 (2014).

\bibitem{KD} D.~V.~Dmitriev, V.~Ya.~Krivnov, Phys.~Rev. B \textbf{92},
184422 (2015).

\bibitem{DKRS} D.~V.~Dmitriev, V.~Ya.~Krivnov, J.~Richter, J.~Schnack,
Phys.~Rev. B \textbf{99}, 094410 (2019); Phys.~Rev. B \textbf{101}, 054427
(2020).

\bibitem{Tonegawa} T.~Tonegawa, M.~Kaburagi, J.~Magn.~Magn.~Matter., \textbf{%
272-276}, 898 (2004).

\bibitem{Kaburagi} M.~Kaburagi, T.~Tonegawa, M.~Kang, J.~Appl.~Phys.,
\textbf{97}, 10B306 (2005).

\bibitem{s12} T.\ Yamaguchi, S.-L.\ Drechsler, Y.\ Ohta, S.\ Nishimoto,
Phys. Rev. B \textbf{101}, 104407 (2020).

\bibitem{ferri} D.~V.~Dmitriev, V.~Ya.~Krivnov, J.\ Phys.: Condens.\ Matter
\textbf{28}, 506002 (2016).

\bibitem{Schnack} O.~Derzhko, J.~Schnack, D.~V.~Dmitriev, V.~Ya.~Krivnov,
J.~ Richter, Eur.~Phys.~J. B \textbf{93}, 161 (2020).

\bibitem{Naturforschung} N.\ Reichert, H.\ Schlueter, T.\ Heitmann, J.\
Richter, R.\ Rausch, and J.\ Schnack, Zeitschrift fuer Naturforschung
\textbf{A}, (2023).

\bibitem{malonate} C.\ Ruiz-Perez, M.\ Hernandez-Molina, P.\ Lorenzo-Luis,
F.\ Lloret, J.\ Cano, M.\ Julve M., Inorg. Chem., \textbf{39}, 3845 (2000).

\bibitem{malonate1} Y.\ Inagaki, Y.\ Narumi, K.\ Kindo, H.\ Kikuchi, T.\
Kamikawa, T.\ Kunimoto, S.\ Okubo, H.\ Ohta, Y.\ Saito, M.\ Azuma, M.\
Takano, H.\ Nojiri, M.\ Kaburagi, T.\ Tonegawa, J. Phys. Soc. Jpn., \textbf{%
74}, 2831 (2005).

\bibitem{InorgChem} C.\ Ruiz-Perez, M.\ Hernandez-Molina, P.\ Lorenzo-Luis,
F.\ Lloret, J.\ Cano, M.\ Julve M., Inorg. Chem., \textbf{39}, 3845 (2000).

\bibitem{Ueda} R.\ Shirakami, H.\ Ueda, H.\ O.\ Jeschke, H.\ Nakano, S.\
Kobayashi, A.\ Matsuo, T.\ Sakai, N.\ Katayama, H.\ Sawa, K.\ Kindo, C.\
Michioka, K.\ Yoshimura, Phys. Rev. B \textbf{100}, 174401 (2019).

\bibitem{F10} A.\ Baniodeh, N.\ Magnani, Y.\ Lan Y., G.\ Buth, C.\ E.\
Anson, J.\ Richter, M.\ Affronte, J.\ Schnack., A.\ K.\ Powell, Npj
Quant.Mater. 2018. V. 3.1. P. 10.

\bibitem{a12} V.~Ya.~Krivnov and D.~V.~Dmitriev, J.~Phys.~Condens.~Matter,
\textbf{35}, 095802 (2023).

\bibitem{s1} D.\ V.\ Dmitriev, V.\ Ya.\ Krivnov, J.\ Phys.: Condens.\ Matter
\textbf{35}, 445802 (2023).

\bibitem{nishimoto} T.\ Yamaguchi, Y.\ Ohta, \& S.\ Nishimoto, Phys.\ Rev.\
B, \textbf{103}, 184410 (2021).

\bibitem{Hida} K.~Hida, Phys.Rev.B\textbf{45}, 2207 (1992).

\bibitem{Hung} H.-H.~Hung, C.-De Gong, Phys.Rev. B\textbf{\ 71}, 054413
(2005).

\bibitem{Sahao} S.~Sahao, D.~Dey, S.K.~Saha, M.~Kumar, J.\ Phys.:Condens.\
Mat. \textbf{32}, 335601 (2020).

\bibitem{Yang} C.\ N.\ Yang and C.\ P.\ Yang, Phys.\ Rev. \textbf{50}, 327
(1966).

\bibitem{Griffits} R.\ B.\ Griffits, Phys.\ Rev.\textbf{133A}, 768 (1964).

\bibitem{Eggert} S.~Eggert, I.~Affleck, M.~Takahashi, Phys.\ Rev.\ Letters
\textbf{73}, 332 (1994).

\bibitem{Klumper} A.~Klumper, D.~C.~Johnson, Phys.\ Rev.\ Letters \textbf{84}%
, 4701 (2000).

\bibitem{Chandra} V.\ Ravi Chandra, D.\ Sen, N.\ B.\ Ivanov, J.\ Richter,
Phys.\ Rev.\ B \textbf{69}, 214406 (2004).

\bibitem{White} S.\ R.\ White, Phys.\ Rev.\ Lett., \textbf{69}, 2863 (1992).

\bibitem{plateau} M.\ Oshikawa, M.\ Yamada, I.\ Affleck, Phys.\ Rev.\
Letters, \textbf{78}, 1997 (1984).

\bibitem{alps} F.\ Alet et al., J.\ Phys.\ Soc.\ Jpn.\ Suppl. \textbf{74},
30 (2005).
\end{thebibliography}
\end{document}